\documentclass[amsmath,amssymb,nofootinbib,superscriptaddress]{revtex4-1}

\usepackage{aas_macros,esint,graphicx}

\usepackage[compact]{titlesec}         
\titlespacing{\subsection}{10pt}{10pt}{10pt}
\titlespacing{\section}{10pt}{10pt}{10pt}

\usepackage[allcolors=blue,colorlinks=true]{hyperref}
\numberwithin{equation}{section}

\makeatletter
\renewcommand{\p@subsection}{}
\renewcommand{\p@subsubsection}{}
\makeatother

\newcommand{\ab}[1]{\left|#1\right|}
\newcommand{\av}[1]{\left\langle#1\right\rangle}
\newcommand{\br}[1]{\left[#1\right]}
\newcommand{\cu}[1]{\left\{#1\right\}}
\newcommand{\pa}[1]{\left(#1\right)}

\newcommand{\dt}{\mathop{}\!\delta}
\newcommand{\ed}{\mathop{}\!\mathrm{d}}
\newcommand{\pd}{\mathop{}\!\partial}

\DeclareMathOperator\sign{sign}
\def\tr{\tilde{r}}
\def\th{\hat{H}}
\def\slr{\mathsf{SL}(2,\mathbb{R})}
\def\slpr{\mathsf{SL}(2,\mathbb{R})_{\rm PR}}
\def\slqn{\mathsf{SL}(2,\mathbb{R})_{\rm QN}}

\begin{document}

\title{\phantom{}\vspace{90pt}\\
\Huge Holography of the Photon Ring
\vspace{30pt}}

\author{\large Shahar Hadar}
\thanks{Department of Mathematics and Physics, University of Haifa at Oranim, Kiryat Tivon 3600600, Israel\\
Haifa Research Center for Theoretical Physics and Astrophysics, University of Haifa, Haifa 3498838, Israel}

\author{\large Daniel Kapec}
\thanks{Center of Mathematical Sciences and Applications, Harvard University, Cambridge, Massachusetts 02138, USA}

\author{\large Alexandru Lupsasca}
\thanks{Princeton Gravity Initiative, Princeton University, Princeton, New Jersey 08544, USA}

\author{\large Andrew Strominger}
\thanks{Center for the Fundamental Laws of Nature, Harvard University, Cambridge, Massachusetts 02138, USA}

\begin{abstract}
\vspace{30pt}
Space-based next-generation interferometers propose to measure the Lyapunov exponents of the nearly bound geodesics that comprise the photon ring surrounding the black hole M87*.
We argue that these classical Lyapunov exponents equal the quantum Ruelle resonances describing the late-time approach to thermal equilibrium of the quantum microstate holographically dual to any Kerr black hole such as M87*.
Moreover, we identify ``near-ring regions'' in the phase space of fields propagating on Kerr that exhibit critical behavior, including emergent conformal symmetries.
These are analogues for sub-extremal Kerr of the much-studied ``near-horizon regions'' of (near-)extremal black holes.
The emergent conformal symmetries greatly constrain the observational predictions for the fine photon ring substructure around M87* and for quasinormal gravitational-wave ringdowns, as well as any proposal for a quantum holographic dual to the Kerr black hole.
More generally, we hope that our identification of several universal features of Kerr spectroscopy provides a useful starting point for a bottom-up approach to holography for astrophysical black holes. 
\end{abstract}

\maketitle

\setcounter{tocdepth}{2}
\tableofcontents

\section{Introduction}

The Bekenstein-Hawking area-entropy law $S_{\rm BH}=\text{Area}/4$ strongly suggests that black holes admit a dual description as a quantum system with $e^{S_{\rm BH}}$ microstates.
A natural question is then: \textit{Where in the bulk geometry do these microstates reside?}
Of course, there may be either zero or several different consistent answers to this question and they may vary depending on the various masses and charges of the black hole, and according to whether the asymptotic region is flat, dS or AdS.
While taking cues from the large body of knowledge about black holes in AdS, we are here primarily interested in astrophysical 
black holes, which are well approximated by the asymptotically flat Kerr metric.

An obvious first guess is that the black hole entropy literally counts all the possible microstates that could lie inside its event horizon.
This guess runs into the ``bag of gold'' problem \cite{Wheeler} and has not been usefully developed thus far.
Another natural guess is that the microstates lie within a few Planck lengths of the horizon.
One of the earliest attempts to reproduce the black hole entropy along these lines \cite{Zurek1985} derived an area law by counting high-frequency but highly redshifted near-horizon modes below a Planckian cutoff.
This resonates with the membrane paradigm \cite{Thorne1986}, in which black hole dynamics is succinctly described by a membrane possessing $\mathcal{O}(1)$ degrees of freedom per Planck area that naturally account for the entropy.

However, there are other indications that the dual quantum state occupies a region that extends over multiple Schwarzschild radii outside the horizon.
In CFT descriptions of black holes with or without string theory, greybody factors needed for the agreement of microscopic and macroscopic scattering are determined by solving differential equations over this larger region \cite{Maldacena1997}.
It is only once these factors have been included that the CFT and semiclassical emission rates agree.
Moreover, recent analyses of information flow using quantum extremal surfaces indicate that infalling bodies are stripped of their quantum information content several Schwarzschild radii prior to reaching the horizon \cite{Penington2020,Almheiri2019,Almheiri2020}.

The question of where the quantum dual resides has taken on renewed interest in light of the unprecedentedly high resolutions recently obtained in black hole imaging \cite{EHT2019a}.
We are now able to directly view the region outside a real black hole in our sky.
The accessible region notably contains the photon ring, which possesses a remarkable subring substructure.
Significantly, this substructure displays highly intricate but universal properties, which follow from general relativity alone and which are insensitive to the detailed composition of the black hole atmosphere \cite{Gralla2019,Johnson2020,Himwich2020,GrallaLupsasca2020a,Hadar2021}.
Wonderfully, this intricate structure is potentially measurable and is a prime target for future space-based VLBI missions \cite{Johnson2020,Gralla2020,GrallaLupsasca2020c,GLM2020,Chael2021}.

In this paper, we argue that the photon ring is indeed a part of the holographic dual for an astrophysical black hole.
Specifically, it encodes Ruelle resonances of the quantum dual (equivalently, quasinormal modes of the black hole) that characterize the late-time decay of a perturbation back to thermal equilibrium.
Intuitively, it is easy to understand this relation.
Black holes are surrounded by photon shells containing the unstably bound orbits whose images produce the photon ring (and its subrings) seen by a distant observer.
Any object thrown at the black hole must cross this photon shell, wherein it generally excites nearly bound photons that leak out of the shell very slowly.
Therefore, the last thing one expects to see as the black hole settles back to its thermal ground state is the photons that have orbited many times before escaping to infinity \cite{Ames1968,Cardoso2021}.\footnote{\label{fn:AdS}In the case of black holes in AdS$_D$ for $D>3$, which we do not consider herein, the photons may bounce around the AdS$_D$ barrier a number of times after leaking off the photon shell, thus only reaching infinity much later and thereby introducing a new time scale.
This obscures the direct relation between the photon ring Lyapunov exponents and their dual Ruelle resonances for the case of AdS spacetimes, and the arguments that we give here are not directly applicable to AdS black holes.
See \cite{Chan1997,Festuccia2008,Cardoso2009} for extensive discussions.}
These trajectories provide eikonal approximations to the quasinormal modes (QNMs) whose damped ringing signals the approach back to equilibrium.
Anticipated measurements of both the photon ring and quasinormal ringdown therefore provide a potentially fertile point of contact between the significant but largely disparate developments over the last several decades in observational and quantum-theoretical black hole physics.

Fully consistent microscopic descriptions of the quantum duals have been given for certain black holes in string theory.
Several largely interrelated proposals have also been made for dual descriptions of astrophysical Kerr black holes such as M87* (see, e.g., \cite{Compere2012}).
However, none of them are complete and the subject remains open and active.
In this paper, we do not assume any specific model.
Rather, our goal is to translate the data from the photon ring into a form that can be used to constrain candidate models for quantum duals to M87*.
Interestingly, our analysis implies a very specific but highly universal form of the \textit{high}-frequency, \textit{short}-distance Ruelle spectrum of the dual theory where the eikonal approximation is valid.
This contrasts with previous macroscopic analyses (which typically probe long-wavelength hydrodynamic properties of the microscopic dual \cite{Bredberg2011,Minwalla2012}) and identifies a second universal regime in the spectrum of quantum systems dual to black holes.
While we refrain from commenting in this paper on how our results bear on proposed duals, we stress that so far no existing proposal has explained the universal structure of the photon ring.
Finding such a universal explanation within a quantum dual is an open challenge for theorists.

A central role in our analysis is played by what we call the \textit{near-ring region}, which is present for any Kerr black hole.
It is an analog of the more familiar near-horizon regions of extremal black hole spacetimes, which are notable for their emergent conformal symmetries.
However, the near-ring region differs in that it is not a region of the {\it spacetime}, but rather of the {\it phase space} of particles or fields propagating on the spacetime geometry.\footnote{Of course, the emergence of conformal symmetry in regions of phase space rather than spacetime is common in condensed matter systems.}
Moreover, it is present for generic black holes, and its existence is not confined to near-extremal ones.
However, the near-ring region does resemble the extremal near-horizon region in that it is characterized by emergent conformal symmetry.

To be more precise, in the special case of a Schwarzschild black hole of mass $M$, the near-ring region for either a massless wave or a photon geodesic with real part of the frequency $\omega_R$ and angular momentum $\ell$ is defined by\footnote{Assuming that the holographic dual lives on a space with time and angular directions, these conditions suggest that the near-ring region is relevant to short distances in the dual theory and that the photon shell itself should perhaps be thought of as the holographic plate.}
\begin{align}
\label{eq:Intro}
    \text{NEAR-RING REGION:}\qquad
    \begin{cases}
        \ab{r-3M}\ll M&\qquad\text{(near-peak)},\\
        \displaystyle\frac{\ell}{\omega_R}-3\sqrt 3M\ll M&\qquad\text{(near-critical)},\vspace{3pt}\\
        \displaystyle\frac{1}{\omega_R}\ll {M}&\qquad\text{(high-frequency)}.
    \end{cases}
\end{align}
Since this definition involves the frequency and angular momentum of the excitations, it pertains to a subregion of phase space, and not of spacetime itself.
The general definition of the Kerr near-ring region is given in section~\ref{sec:Kerr} below.

We show herein that photon trajectories in the near-ring region are acted on by an emergent\footnote{$\slpr$ actually admits an extension as a group action over the entire phase space, but it appears to be of practical utility only in the near-ring region where the expressions become very simple and local.} conformal symmetry group which we refer to as $\slpr$.
Geodesics for which the LHS of the first two inequalities in \eqref{eq:Intro} are strictly zero are bound orbits comprising the codimension-2 submanifold of the (affinely parameterized) null geodesic phase space known as the photon shell.
This shell is an invariant subspace of the $\slpr$ action, (i.e., it is a fixed point), and the near-ring region is its near-neighborhood.
There exists a scaling subgroup of $\slpr$ that drives all photon geodesics to the fixed point.
Near-ring photons that are not exactly on the fixed point can orbit the black hole multiple times before escaping and potentially arriving at a distant telescope.
Such photons form an important part of the black hole image known as the photon ring, whose observation is proposed in \cite{Johnson2020,GLM2020}.
We show that the successive subrings of the photon ring, labelled by orbit number, transform into one another under a discrete scaling subgroup of $\slpr$ that preserves the position of the observer's telescope.
This discrete subgroup is generated by finite scaling transformations $e^{-\gamma H_0}$, where $H_0$ is the $\slpr$ dilation generator and $\gamma$ a Lyapunov exponent controlling the demagnification of successive subring images.
Hence, a future measurement of these Lyapunov exponents would constitute a detection of a qualitatively new kind of emergent symmetry of nature.
We hope that this will serve as a further incentive to the ongoing efforts to measure the fine structure of the photon ring. 

It is also of interest to consider massless wave propagation in the near-ring region \eqref{eq:Intro}.
It is well-known that in this approximation, the massless wave equation reduces to the Schr\"{o}dinger equation for the upside-down harmonic oscillator.
This system has a well-known conformal symmetry\footnote{Arising as a square of the harmonic oscillator algebra.} which we denote $\slqn$, and the eikonal QNMs fall into highest-weight representations of this symmetry algebra.
Our construction is highly reminiscent of the observations made for unwarped \cite{Birmingham2002} and warped \cite{Chen2009} BTZ black holes that quasinormal modes form representations of the AdS$_3$ conformal group.
This hints that $\slqn$ may actually act on the full set of quantum black hole microstates, although we certainly do not establish any such concrete connection here. 

Solutions to the massless wave equation are related, in the eikonal (geometric optics) approximation, to congruences of null geodesics.
The eikonal approximation is valid in the subset of the near-ring region where $\omega_R(r-3M)^2\gg M$.
We show explicitly that the tower of quasinormal modes can be constructed from the so-called homoclinic geodesics, for which $\frac{\ell}{\omega_R}=3\sqrt 3M$ exactly, and we provide a simple new geometric derivation of the overtone wavefunctions that applies to all black hole spacetimes and that follows straightforwardly from properties of the photon ring.

We have found two\footnote{We will also encounter a third conformal group of potential interest, denoted $\widehat{\mathsf{SL}}(2,\mathbb{R})$ and discussed in section \ref{sec:Schwarzschild}, which is more closely tied to isometries.} conformal groups, $\slpr$ and $\slqn$, both of which emerge only in the near-ring region.
$\slpr$ acts naturally on geodesics, while $\slqn$ acts naturally on quasinormal modes.
They clearly bear some relation to each other, but from what we have understood so far the precise connection is rather subtle, and we leave this question unanswered for now. 
Conformal symmetries have played a central role in the microscopic accounting for the Bekenstein-Hawking black hole entropy \cite{Strominger1996,Strominger1998,Guica2009}.
It is natural to ask if the near-ring emergent conformal symmetries described here could potentially also play such a role.
We'd like to know the answer! 

One might ask why we are so interested in the small and peculiar near-ring region of a black hole.
The answer for observers is that this is the region that dominates the portion of the black hole image belonging to the black hole itself and not to the circulating plasma.
The answer for pure theorists is that the emergent symmetries in this region may provide important clues for the construction of the holographic duals of real-world black holes.
More generally, regions with emergent symmetries are almost always of special interest. 

We hope to have made some observations that will constrain and provide a jumping-off point for future attempts at a bottom-up construction of the holographic duals of astrophysical black holes, and to have deepened the theoretical understanding of the structure being probed by recent and continuing spectacular advances in observational black hole astrophysics.

The outline of the remainder of this paper is as follows.
Section~\ref{sec:Schwarzschild} contains an analysis of the Schwarzschild photon ring, its relation to the eikonal quasinormal mode spectrum, and the emergence of $\slpr$ and $\slqn$ symmetries in the near-ring region.
The Kerr analysis is technically more complicated but conceptually identical and appears in section~\ref{sec:Kerr}.
We close in section~\ref{sec:KerrHologram} with a discussion of the relation between classical photon ring Lyapunov exponents of the Kerr geometry and quantum Ruelle exponents of the purported holographic dual.

\section{The Schwarzschild black hole}
\label{sec:Schwarzschild}

In this section, we consider the photon ring and quasinormal mode spectrum of the four-dimensional Schwarzschild black hole.
In coordinates $(t,r,x^a)$ with $x^a=(\theta,\phi)$, its line element is
\begin{align}
	ds^2=-\pa{1-\frac{2M}{r}}\ed t^2+\pa{1-\frac{2M}{r}}^{-1}\ed r^2+r^2\gamma_{ab}\ed x^a\ed x^b\;,
\end{align}
where $\gamma_{ab}\ed x^a\ed x^b=\ed\theta^2+\sin^2{\theta}\ed\phi^2$ is the round metric on the sphere and $f(r)=1-\frac{2M}{r}$ is the blackening factor.

\subsection{The near-ring region}

The Schwarzschild geometry is static and spherically symmetric, so geodesic motion is confined to lie in a plane that we will take to be equatorial; all other trajectories can be obtained by symmetry transformations.
For a null geodesic $(t(s),r(s),x^a(s))$ parameterized by affine time $s$,
\begin{align}
    -f(r)\pa{\frac{dt}{ds}}^2+\frac{1}{f(r)}\pa{\frac{dr}{ds}}^2+r^2\gamma_{ab}\frac{dx^a}{ds}\frac{dx^b}{ds}=0\;.
\end{align}
Throughout this paper, we always consider affinely parameterized geodesics.
This associates an energy (or frequency) to a photon traveling along such a geodesic. 
The energy and angular momentum in the equatorial plane are
\begin{align}
    E=f(r)\frac{dt}{ds}\;,\qquad
    L=r^2\frac{d\phi}{ds}\;,
\end{align}
so the null geodesic equation takes the form 
\begin{align}
    \label{eq:SchGeo}
    -\pa{\frac{dr}{ds}}^2+\mathcal{V}(r)=0\;,\qquad
    \mathcal{V}(r)=E^2-f(r)\frac{L^2}{r^2}\;.  
\end{align}
We will only consider geodesics with $L>0$.
Those with $L<0$ can be obtained by a rotation.

Spherical photon orbits require $\mathcal{V}(r)=\mathcal{V}'(r)=0$.
The second condition reads
\begin{align}
    \mathcal{V}'(r)=-f'(r)\frac{L^2}{r^2}+f(r)\frac{2L^2}{r^3}
    =0\;,
\end{align}
so such an orbit can only exist at the critical orbital radius
\begin{align}
    \label{eq:SchCriRad}
    \tr=3M\;.
\end{align}
The condition $\mathcal{V}(\tr)=0$ then requires the energy-rescaled angular momentum $\lambda=\frac{L}{E}$ to take the critical value
\begin{align}
    \label{eq:SchCriMom}
    \tilde{\lambda}=3\sqrt{3}M\;.
\end{align}
The angular velocity $\tilde{\Omega}$ and orbital half-period $\tau$ of the bound photon orbit are therefore
\begin{align}
    \label{eq:SchOrb}
    \tilde{\Omega}=\frac{1}{3\sqrt{3}M}\;,\qquad
    \tau=3\sqrt{3} M\pi\;.
\end{align}
We will consider nearly bound null geodesics with small radial deviation $\dt r=r-\tr$.
More specifically, we are interested in the near-ring region defined \textit{in phase space} by
\begin{align}
    \label{eq:SchGeoNearRing}
    \text{NEAR-RING REGION:}\qquad
    \begin{cases}
        \ab{\dt r}\ll M&\qquad\text{(near-peak)}\;,\\
        |\lambda-\tilde{\lambda}|\ll M&\qquad\text{(near-critical)}\;,\\
        \displaystyle\frac{1}{E}\ll M&\qquad\text{(high-energy)}\;.
    \end{cases}
\end{align}
The first condition zooms in on the bound orbit in spacetime, while the second condition zooms in on the bound orbit in momentum space.
Taken together, these conditions scale into the region of phase space known as the photon shell, defined as the locus $\dt r=0=\lambda-\tilde{\lambda}$.\footnote{This is distinct from the photon ring itself, which is usually defined as the ring image produced by near-shell photons when they reach a telescope at infinity.
Sometimes, however, the term ``photon ring'' is used with a more general meaning.}
The last condition is required in order to relate solutions of the wave equation to geodesic congruences and will only become important in the discussion of quasinormal modes in the next section.

Linearizing about the near-ring region, one finds that
\begin{align}
    \frac{d\dt r}{ds}=\sqrt{\mathcal{V}(\tr+\dt r)}
    \approx\sqrt{\frac{1}{2}\mathcal{V}''(\tr)}\dt r
    =\frac{L}{(3M)^2}\dt r\;,
\end{align}
which implies that slightly perturbed near-critical orbits diverge exponentially in coordinate time:
\begin{align}
    \frac{d\dt r}{dt}\approx\frac{\dt r}{3\sqrt{3}M}\;.
\end{align}
The Lyapunov exponent is therefore
\begin{align}
    \label{eq:SchLyp}
    \dt r(t)\approx e^{\gamma_L t}\dt r_0\;,\qquad
    \gamma_L=\frac{1}{3\sqrt{3}M}\;.
\end{align}
Together with $\phi\approx\tilde{\Omega}t$, this gives the solution of the Schwarzschild null geodesic equation in the near-ring region \eqref{eq:SchGeoNearRing}.

We have derived the explicit form of the geodesics only in the near-ring region.
A generic geodesic will leave this region in finite time.
The full solution is of course more complicated, but given the angular momentum and energy, it is fully determined by an ODE in the radial variable $r$.
Most of the interesting motion occurs in the near-ring region.

\subsection{Conformal symmetry of the quasinormal mode spectrum}

In this subsection, we solve for the eikonal quasinormal modes in the Schwarzschild near-ring region and show that they form a shadow pair of highest-weight representations of the emergent near-ring $\slqn$ symmetry. 

In the linearized approximation to black hole dynamics, scalar perturbations obey the wave equation\footnote{We could also consider the moderately more complicated case of photons and gravitons, but the effects of spin are subleading in the eikonal and near-ring limits of interest to us here, so \eqref{eq:Wav} suffices for our purposes.}
\begin{align}
    \label{eq:Wav}
    \nabla^2\Phi(x)=0\;.
\end{align}
The Schwarzschild black hole has a canonically normalized Killing vector $\pd_t$ that we use to define frequencies.
Spherical symmetry allows for a mode decomposition in terms of spherical harmonics $Y_{\ell m}(\theta,\phi)$:
\begin{align}
    \label{eq:SchAns}
    \Phi(t,r,\theta,\phi)=\int\ed\omega\sum_{\ell=0}^\infty\sum_{m=-\ell}^{\ell}c_{\ell m}(\omega)\Phi_{\ell m\omega}(t,r,\theta,\phi)\;,\qquad
    \Phi_{\ell m\omega}(t,r,\theta,\phi)=e^{-i\omega t}\frac{\psi_{\ell\omega}(r)}{r}Y_{\ell m}(\theta,\phi)\;.
\end{align}
The radial part of the wave equation takes the form
\begin{align}
    \label{eq:SchRad}
    \br{\pd_{r_*}^2+V(r_*)}\psi_{\ell\omega}(r_*)=0\;,
\end{align}
where $r_*=r+2M\log\pa{\frac{r}{2M}-1}$ is the tortoise coordinate and the wave potential is given by
\begin{align}
    \label{eq:SchWavPot}
    V(r_*)=\omega^2-f(r)\br{\frac{\ell(\ell+1)}{r^2}+\frac{2M}{r^3}}\;.
\end{align}
Quasinormal modes correspond to solutions of \eqref{eq:SchRad} that obey an ingoing boundary condition $\Phi\sim e^{-i\omega r_*}$ at the horizon ($r_*\to-\infty$) and an outgoing boundary condition $\Phi\sim e^{i\omega r_*}$ at spatial infinity ($r_*\to\infty$).
The imposition of two boundary conditions on this second-order ODE defines a ``shooting problem'' that results in a discrete spectrum
\begin{align}
    \omega=\omega_R+i\omega_I\;,
\end{align}
which is in general complex due to the non-Hermitian (dissipative) boundary condition.
The associated solutions are interpreted as short-lived resonances whose lifetimes are typically set by the temperature of the black hole.
They come in families $\psi_{\ell n}(r)$ with discrete frequencies $\omega_{\ell n}$ labelled by an overtone number $n$, with higher overtones decaying exponentially faster.
Exact solutions to this problem are rare (although exceptions exist in spacetimes with $\slr$ isometries) and one is typically forced to resort to numerical approximation schemes.
However, there is a ``near-ring'' limit that can be understood analytically and that connects directly with the geometry of the photon shell and its (nearly) bound geodesics reviewed in the previous section.

When $\omega_R$ and $\ell$ are both large and of comparable magnitude, the wave potential \eqref{eq:SchWavPot} may be approximated by
\begin{align}
    V(r)\approx\omega_R^2-f(r)\frac{\ell^2}{r^2}\;,
\end{align}
which matches the geodesic potential \eqref{eq:SchGeo} provided that we identify $E=\omega_R$ and $L=\ell$.
This observation is the basis for the geometric optics approximation discussed in the next section, and applies everywhere in the spacetime.

Here we want to solve the wave equation in the near-ring region, which we define in analogy to \eqref{eq:SchGeoNearRing} as
\begin{align}
    \label{eq:SchWavNearRing}
    \text{NEAR-RING REGION:}\qquad
    \begin{cases}
        \ab{\dt r}\ll M&\qquad\text{(near-peak)}\;,\\
        \displaystyle\ab{\frac{\ell}{\omega_R}-\tilde{\lambda}}\ll M&\qquad\text{(near-critical)}\;,\vspace{3pt}\\
        \displaystyle\frac{1}{\omega_R}\ll M&\qquad\text{(high-frequency)}\;.
    \end{cases}
\end{align}
This defines a region of the phase space of waves on Schwarzschild rather than simply a region of spacetime. 
The wave potential \eqref{eq:SchWavPot} in this near-ring region is
\begin{align}
    V(\dt r)\approx\frac{\omega_R^2}{3M^2}\dt r^2+2i\omega_R\omega_I\;,
\end{align}
and the radial ODE \eqref{eq:SchRad} reduces to
\begin{align}
    \br{\pd_{r_*}^2+\frac{\omega_R^2}{3M^2}\dt r^2+2i\omega_R\omega_I}\psi(\dt r)=0\;.
\end{align}
Near the critical radius $\tr=3M$, $\pd_{r_*}\approx\tilde{f}\pd_{\dt r}$ with $\tilde{f}=1-\frac{2M}{\tr}=\frac{1}{3}$, so we may rewrite this as
\begin{align}
    \mathcal{H}\psi=i\omega_I\psi\;,\qquad
    \mathcal{H}=-\frac{1}{2\omega_R}\br{\pd_x^2+\gamma_L^2\omega_R^2x^2}\;,\qquad
    x=r_*-\tr_*
    =\frac{\dt r}{\tilde{f}}\;,
\end{align}
which we recognize as the time-independent Schr\"odinger equation for eigenstates $\psi$ of the inverted harmonic oscillator with associated eigenvalues $i\omega_I$.
The eigenvalues are imaginary because the boundary conditions are non-Hermitian. 
Following \cite{Subramanyan2021,Raffaelli2022}, we now define the operators 
\begin{align}
    \label{eq:SLR}
    a_\pm=\frac{e^{\pm\gamma_Lt}}{\sqrt{2\gamma_L\omega_R}}\pa{\mp i\pd_x-\gamma_L\omega_Rx}\;,\qquad
    L_0&=-\frac{i}{4}\pa{a_+a_-+a_-a_+}
	=\frac{i}{2\gamma_L}\mathcal{H}\;,\qquad
	L_\pm=\pm\frac{a_\pm^2}{2}\;.
\end{align}
The $a_\pm$ generate the Heisenberg algebra $\br{a_+,a_-}=iI$, while the $L_m$ obey the exact $\slqn$ commutation relations:
\begin{align}
    \br{L_0,L_\pm}=\mp L_\pm\;,\qquad
    \br{L_+,L_-}=2L_0\;.
\end{align}
These operators are defined everywhere but are of interest only in the near-ring region where $L_0$ is proportional to the Hamiltonian. 
Eigenstates of $L_0$ satisfy $L_0\psi_h=h\psi_h$,
so we identify $\omega_I=-2\gamma_Lh$.
The mode ansatz \eqref{eq:SchAns} reduces in the near-ring region \eqref{eq:SchWavNearRing} to
\begin{align}
    \Phi_{\ell m\omega}(t,r,\theta,\phi)\approx e^{-i\omega_Rt}\frac{\Phi_h(t,x)}{r}Y_{\ell m}(\theta,\phi)\;,\qquad
    \Phi_h(t,x)=e^{\omega_It}\psi_h(x)
    =e^{-2\gamma_Lht}\psi_h(x)\;.
\end{align}
The operator $L_0$ obviously has no normalizable ground state, but it can still have a discrete spectrum if the boundary conditions are chosen appropriately.
The quasinormal mode boundary condition is equivalent to the imposition of a highest-weight condition $L_+\Phi_h=0$ on the fundamental mode.
There are two solutions with $h=\frac{1}{4}$ and $h=\frac{3}{4}$:
\begin{align}
    \Phi_\frac{1}{4}(t,x)&=e^{-\frac{1}{2}\gamma_Lt}\psi_\frac{1}{4}(x)\;,
    &&\psi_\frac{1}{4}(x)=e^{\frac{i}{2}\gamma_L\omega_Rx^2}\;,\\
    \Phi_\frac{3}{4}(t,x)&=e^{-\pa{1+\frac{1}{2}}\gamma_Lt}\psi_\frac{3}{4}(x)\;,
    &&\psi_\frac{3}{4}(x)=xe^{\frac{i}{2}\gamma_L\omega_Rx^2}\;.
\end{align}
Higher overtones are then obtained as $\slqn$-descendants:
\begin{align}
    \Phi_{h,N}(t,x)=L_-^N\Phi_h(t,x)
    =e^{-2\gamma_L(h+N)t}\psi_{h+N}(x)
    \propto e^{-2\gamma_L(h+N)t}D_{2(h+N)-\frac{1}{2}}\pa{\sqrt{-2i\gamma_L\omega_R}x}\;,
\end{align}
where $D_n(x)$ denotes the $n^\text{th}$ parabolic cylinder function.

The two towers of $\slqn$-descendants obtained from the primary states with $h=\frac{1}{4}$ and $h=\frac{3}{4}$ are the QNM overtones in the near-ring region \eqref{eq:SchWavNearRing}, with the $n^\text{th}$ overtone corresponding to the state with
\begin{align}
    N=\frac{1}{2}\pa{n+\frac{1}{2}}-h\;,\qquad
    h=\begin{cases}
        \frac{1}{4}&\qquad\text{if $n$ is even}\; ,\\
        \frac{3}{4}&\qquad\text{if $n$ is odd}\; .
    \end{cases}
\end{align}
In other words, the QNM overtones fall into two irreps of the $\slr$ generated by \eqref{eq:SLR}.
The Casimir is
\begin{align}
    \mathcal{C}\Phi_{h,N}\equiv\pa{-L_0^2+\frac{L_+L_-+L_-L_+}{2}}\Phi_{h,N}
    =h(1-h)\Phi_{h,N}\; ,
\end{align}
so the two representations that appear are shadows of each other with Casimir
\begin{align}
    \mathcal{C}=\frac{3}{16}\; .
\end{align}
These two irreps combine to form a single irrep of the Heisenberg algebra generated by $a_\pm$ and $I$. 

Algebraically, it is also possible to trivially combine these two representations into a single highest-weight $h=\frac{1}{2}$ representation of a different $\widehat{\mathsf{SL}}(2,\mathbb{R})$.
One defines
\begin{align}
    \hat{L}_0=2L_0\;,
\end{align}
along with 
\begin{align}
    \hat{\Phi}_{\frac{1}{2}+n}=
    \begin{cases}
        \Phi_{\frac{1}{4}+\frac{n}{2}}&\qquad\text{if $n$ is even}\;,\\
        \Phi_{\frac{3}{4}+\frac{n-1}{2}}&\qquad\text{if $n$ is odd}\;.
    \end{cases}
\end{align}
The action of all three $\hat{L}_m$ is then simply defined by the commutation relations $[\hat{L}_m,\hat{L}_n]=(m-n)\hat{L}_{m+n}$.
However, with the exception of $\hat{L}_0$, we have so far been unable to explicitly represent this $\widehat{\mathsf{SL}}(2,\mathbb{R})$ in terms of a differential operator or an action on phase space.
We leave this to future work.

We note that at the edges $\dt r\to\pm\infty$ of the near-peak region,
\begin{align}
    D_n\pa{\sqrt{-2i\gamma_L\omega_R}x}\stackrel{x\to\pm\infty}{\sim}x^ne^{\frac{i}{2}\gamma_L\omega_Rx^2}\;.
\end{align}
Therefore, the $n^\text{th}$ overtone behaves near the edges as
\begin{align}
    \label{eq:SchOvt}
    \Phi_{\ell mn}(t,r,\theta,\phi)\stackrel{x\to\pm\infty}{\sim}e^{-\pa{n+\frac{1}{2}}\gamma_L t}x^ne^{-i\omega_R\pa{t-\frac{1}{2}\gamma_Lx^2}}Y_{\ell m}(\theta,\phi)\;,\qquad
    \omega_R=\frac{\ell+\frac{1}{2}}{\tilde{\lambda}}\;.
\end{align}
Recalling that $\tilde{\Omega}=\frac{1}{\tilde{\lambda}}=\frac{1}{3\sqrt{3}M}=\gamma_L$, the eikonal QNM spectrum of a Schwarzschild black hole is
\begin{align}
    \omega_{\ell n}=\pa{\ell+\frac{1}{2}}\tilde{\Omega}-i\pa{n+\frac{1}{2}}\gamma_L \;.
\end{align}

We have given the analytic form of the QNMs only in the near-ring region \eqref{eq:SchWavNearRing}.
The full solutions with the same conserved quantities $(\omega_R,\omega_I,\ell,m)$ can be extended everywhere in the spacetime by solving the exact radial ODE \eqref{eq:SchRad}.
The full wavefunctions still obey QNM boundary conditions since the photon shell is the only place where the radial momentum can change sign.
Outside the near-ring region, these solutions still form $\slqn$ multiplets but they are no longer related by the simple operators given in \eqref{eq:SLR}, which fail to commute with the wave equation and do not map solutions to solutions.
Although not present globally, the $\slqn$ symmetry of the wave equation in the near-ring region  is enough to derive a conformal multiplet structure for the full spacetime QNM solutions.

\subsection{Quasinormal modes from geometric optics}
\label{sec:QNMGeo}

In the eikonal limit of large frequencies, the geometric optics approximation relates solutions of the massless wave equation to null geodesic congruences.
Applied to the null congruences in the neighborhood of the photon shell---the ``eikonal near-ring'' region---this approximation has been shown in a variety of circumstances \cite{Goebel1972,Ferrari1984,Mashhoon1985,Iyer1987a,Iyer1987b,Seidel1990,Decanini2003,Dolan2009,Cardoso2009,Dolan2010,Dolan2011,Decanini2010,Yang2012} to reproduce the eikonal QNM frequencies, with the real and imaginary parts respectively given in terms of the orbital frequency and Lyapunov exponents of (nearly) bound orbits.
We will see below that the region of validity of the eikonal approximation overlaps with a subset of the near-ring region.

In this short-wavelength, large-$\omega$ limit, one approximates a solution to the wave equation \eqref{eq:Wav} in terms of a rapidly oscillating phase $S(x)$ and a slowly varying amplitude $A(x)$:
\begin{align}
    \Phi(x)\sim A(x)e^{iS(x)}\;.
\end{align}
In terms of the gradient of the phase
\begin{align}
    p_\mu=\pd_\mu S(x)\;,
\end{align}
the wave equation takes the form
\begin{align}
    \label{eq:ShortWave}
    -p_\mu p^\mu A+i\pa{2p^\mu\nabla_\mu A+\nabla_\mu p^\mu A}+\nabla^2A=0\;,
\end{align}
and one attempts to solve this equation order by order in $p_\mu$.
The leading-order term implies that $p_\mu$ is a null vector satisfying the geodesic equation:
\begin{align}
    p_\mu p^\mu=0\;,\qquad
    p^\mu\nabla_\mu p_\nu=0\;.
\end{align}
The second equality follows from the first since $p_\mu$ is a gradient, and it implies that $s$ defined by \begin{align}
    \pd_s=p^\mu\pd_\mu
\end{align}
is an affine parameter.
The first equation also implies that the phase $S(x)$ is a solution of the Hamilton-Jacobi equation for the covariant geodesic Hamiltonian $\frac{1}{2}g^{\mu\nu}p_\mu p_\nu$.

The subleading term in \eqref{eq:ShortWave} relates the expansion $\theta=\nabla_\mu p^\mu$ of the null congruence to the directional derivative of the wave amplitude along the geodesic:
\begin{align}
    \label{eq:GeoAmp}
    p^\mu\pd_\mu\log{A(x)}=-\frac{1}{2}\theta(x)\;.
\end{align}
For a congruence with positive expansion, the amplitude of the wave therefore decays exponentially with affine time:
\begin{align}
    \label{eq:AmpDec}
    \pd_s\log{A(x)}=-\frac{1}{2}\theta(x)
    \qquad\implies\qquad
    A\sim A_0e^{-\frac{1}{2}\theta s}\;.
\end{align}
To summarize, given a (rotation-free) null congruence with local tangent $p^\mu(x)$ and expansion $\theta(x)$, one can construct an approximate solution to the wave equation whose wavefronts (level sets of constant phase $S$) propagate along null geodesics.
Quasinormal modes correspond to trajectories that are asymptotically bound to the photon shell, and the exponential divergence of rays near the shell determines the exponential decay of the quasinormal modes with time.

In the case of null equatorial Schwarzschild geodesics, the Hamilton-Jacobi principal function is
\begin{align}
    S(t,r_*,\phi)=-Et+L\phi\pm\int^{r_*}\sqrt{\mathcal{V}(r_*')}\ed r_*'\;.
\end{align}
For a null congruence to satisfy the quasinormal mode boundary conditions, the potential $\mathcal{V}(r_*)$ must have a double root, i.e., the geodesics must have momentum satisfying the critical condition \eqref{eq:SchCriMom} and be asymptotically bound to the photon shell.
Evaluating the radial integral from the critical radius \eqref{eq:SchCriRad} then gives
\begin{align}
    \tilde{S}(t,\dt r,\phi)=E\pa{-t+\tilde{\lambda}\phi+\int_0^{\dt r}\sqrt{\frac{\pa{9M+\dt r'}\dt r'^2}{\pa{3M+\dt r'}\pa{M+\dt r'}^2}}\ed\dt r'}
    \stackrel{\ab{\dt r}\to 0}{\approx}E\pa{-t+\tilde{\lambda}\phi+\frac{\sqrt{3}}{2M}\dt r^2}\;.
\end{align}
Noting that $\frac{\sqrt{3}}{2M}\dt r^2=\frac{1}{2}\gamma_Lx^2$ and identifying $E=\omega_R$ and $L=m$ as in the previous section, we recognize this as the phase of \eqref{eq:SchOvt} with $m=+\ell$.
The critical geodesics have an expansion
\begin{align}
    \tilde{\theta}=\nabla^2\tilde{S}
    \stackrel{\ab{\dt r}\to 0}{\approx}3\gamma_L\omega_R\,,
\end{align}
and hence, a simple solution describing the wave amplitude \eqref{eq:GeoAmp} of the QNM null congruence is $A_0(t,\dt r,\phi)=e^{-\frac{1}{2}\gamma_Lt}$.
The eikonal approximation is valid as long as $\gamma_L\omega_R\dt r^2\gg 1$, which for Schwarzschild amounts to the condition $\omega_R\dt r^2\gg M$.
As such, the eikonal region overlaps with the near-ring region \eqref{eq:SchWavNearRing} when $\frac{M}{\omega_R}\ll\dt r^2\ll M^2$.

According to \eqref{eq:AmpDec}, any quantity which is constant along the null congruence can be used to produce new solutions to the wave equation from a given seed solution.
In other words, if a function $u(x)$ satisfies 
\begin{align}
    \label{eq:GeoCon}
    p^\mu\pd_\mu u=0\;,
\end{align}
and moreover if
\begin{align}
    \label{eq:A0}
    \Phi_0(x)=A_0(x)e^{iS(x)}
\end{align}
is a solution to \eqref{eq:ShortWave}, then
\begin{align}
    \Phi_n(x)=u^n(x)A_0(x)e^{iS(x)}
\end{align}
is also an approximate solution to the wave equation.
In all the examples that we consider (including the one above), the minimal solution to the amplitude equation \eqref{eq:GeoAmp} takes the form $A_0\sim e^{-\frac{1}{2}\gamma_Lt}$, and
a straightforward analysis of the geodesic equation indicates that the quantity
\begin{align}
    u(x)=e^{-\gamma_Lt}\dt r
\end{align}
is constant on the unstable homoclinic orbit, and therefore obeys \eqref{eq:GeoCon}.
Hence, we can immediately write down a family of solutions associated to the same critical orbit with phase $\tilde{S}$, but differing in the amplitude:
\begin{align}
    \Phi_n(x)\sim e^{-\pa{n+\frac{1}{2}}\gamma_L t}\dt r^ne^{i\tilde{S}(x)}\;.
\end{align}
This eikonal QNM approximation agrees with the near-ring approximation \eqref{eq:SchOvt} in their overlap region.

\subsection{Observable conformal symmetry of the photon ring}
\label{subsec:SchScr}

In this subsection, we identify another emergent near-ring conformal symmetry, denoted $\slpr$, which acts on null geodesics rather than waves.
A discrete subgroup of this scaling symmetry preserves the endpoints of null geodesics terminating at a fixed telescope and maps successive subrings to one another in black hole images.
This emergent scaling symmetry is potentially observable with upcoming space-based VLBI missions.

As before, spherical symmetry allows us to restrict our attention to geodesics in the equatorial plane.
Let $\Gamma$ denote the four-dimensional phase space of colored, equatorial null geodesics in Schwarzschild with coordinates $(r,\phi,p_r,p_\phi)$ and canonical symplectic form.
Time evolution is generated by the Hamiltonian
\begin{align}
	\label{eq:SchH}
	H(r,p_r,p_\phi)=\sqrt{\pa{1-\frac{2M}{r}}\br{\frac{p_\phi^2}{r^2}+\pa{1-\frac{2M}{r}}p_r^2}}\;,
\end{align}
which is obtained by solving the null condition $g^{\mu\nu}p_\mu p_\nu=0$ for $p_t=-H$.
Inverting \eqref{eq:SchH} gives
\begin{align}
	\label{eq:SchPr}
	p_r(r,H,L)=\pm\frac{\sqrt{\mathcal{V}(r)}}{f(r)}\;,\qquad \mathcal{V}(r)=H^2-f(r)\frac{L^2}{r^2}\;,
\end{align}
where $L=p_\phi$.
The coordinate transformation $(r,\phi,p_r,p_\phi)\to(T,\Phi,H,L)$ defined by 
\begin{align}
	\ed T&=\frac{H}{f(r)\sqrt{\mathcal{V}(r)}}\ed r\;,\qquad
	\ed\Phi=\ed\phi-\frac{L}{r^2\sqrt{\mathcal{V}(r)}}\ed r\;, 
\end{align}
is canonical since it preserves the symplectic form
\begin{align}
	\Omega=\ed p_r\wedge\ed r+\ed p_\phi\wedge\ed\phi
	=\ed H\wedge\ed T+\ed L\wedge\ed\Phi\;.
\end{align}
These action-angle variables lead to trivial equations of motion:
\begin{align}
	\dot{H}=\cu{H,H}=0\;,&&
	\dot{L}=\cu{L,H}=0\;,&&
	\dot{\Phi}=\cu{\Phi,H}=0\;,&&
	\dot{T}=\cu{T,H}=1\;.
\end{align}
The first two equations indicate that the phase space $\Gamma$ foliates into superselection sectors of fixed $(H,L)$, which are conserved momenta.
The third equation implies that the Hamiltonian flow sends a photon with initial coordinates $(r_s,\phi_s,H,L)$ to final coordinates $(r_o,\phi_o,H,L)$ according to the rule
\begin{align}
	\label{eq:Azimuth}
	\Delta\phi=\phi_o-\phi_s
	=\fint_{\phi_s}^{\phi_o}\ed\phi
	=\fint_{r_s}^{r_o}\frac{L}{r^2\sqrt{\mathcal{V}(r)}}\ed r\;,
\end{align}
where the slash indicates that the integral is to be evaluated along the photon trajectory.
The last equation identifies $T$ as the variable conjugate to energy, i.e., time.
Hence, the time elapsed during evolution from a state $(r_s,\phi_s,H,L)$ to $(r_o,\phi_o,H,L)$ is
\begin{align}
	\label{eq:TimeLapse}
	T=\fint_{r_s}^{r_o}\frac{H}{f(r)\sqrt{\mathcal{V}(r)}}\ed r\; .
\end{align}
Equations \eqref{eq:Azimuth} and \eqref{eq:TimeLapse} are the usual solution to the null geodesic equation in Schwarzschild.
These integrals can be evaluated explicitly in terms of elliptic functions \cite{GrallaLupsasca2020b}, but their detailed form will not be needed here.
The important point to note is that the integral \eqref{eq:TimeLapse} diverges logarithmically when $\mathcal{V}(r)$ has a double root.
As such, the time function therefore diverges for a homoclinic trajectory with one endpoint on the photon shell, and therefore $T$ is only a local coordinate on phase space. 

Since we are concerned with optical images, we focus on geodesics that begin and end at null infinity, always remaining outside the sphere of bound photon orbits at $\tr=3M$.
These have
\begin{align}
    \label{eq:SchHhat}
	\th\equiv H-\frac{\ab{L}}{3\sqrt{3}M}<0\;.
\end{align}
A distant observer at large radius $r_o\to\infty$ receives these geodesics with impact parameter
\begin{align}
	\label{eq:ImpactParameter}
	b=\frac{\ab{L}}{H}>3\sqrt{3}M\;.
\end{align}
Their radius of closest approach is reached when the radial momentum \eqref{eq:SchPr} vanishes.
This occurs at the largest root of the radial potential $\mathcal{V}(r)$ \cite{Gates2020}: 
\begin{align}
	\label{eq:RadialTurningPoint}
	r_{\rm min}=\frac{2b}{\sqrt{3}}\cos\br{\frac{1}{3}\arccos\pa{-\frac{3\sqrt{3}M}{b}}}
	>3M\;.
\end{align}
Geodesics with $\th=0$ are homoclinic and asymptote to the closed photon orbit at $\tr=3M$ in the far past and/or future.
Their impact parameter $b=3\sqrt{3}M$ defines the critical curve in the observer sky.

Using the fact that the coordinates $(T,\Phi,H,L)$ are canonical, it is straightforward to check that the functions
\begin{align}
	\label{eq:Generators}
	H_+=\th\;,\qquad
	H_0=-\th T\;,\qquad
	H_{-}=\th T^2\;,
\end{align}
obey the $\slpr$ algebra.
This conformal algebra commutes with $L=p_\phi$ and therefore acts within superselection sectors $\Gamma_L$ of fixed angular momentum.
However, as indicated by \eqref{eq:Generators}, it does modify the energy (or photon color) $H=\th+\frac{\ab{L}}{3\sqrt{3}M}$.
Thus, it will also modify the impact parameter \eqref{eq:ImpactParameter} as well as the radius of closest approach \eqref{eq:RadialTurningPoint}.
The $\slpr$ group action on the unbound elements of $\Gamma_L$ is transitive: finite $\slpr$ transformations can be used to map any such geodesic to any other.
Of course, one could replace $\th$ in this construction with any function of the form $H-g(L)$, but only the choice \eqref{eq:SchHhat} leads to dilations that scale into the photon shell as in \eqref{eq:FiniteDilation} below.

We are particularly interested in $\slpr$-invariant submanifolds of phase space, which are sets of points $(\tr,\tilde{\phi},\tilde{p}_r,\tilde{p}_\phi)$ left invariant by all three $\slpr$ generators: 
\begin{align}
	\left.\cu{H_m,x}\right|_{(\tr,\tilde{\phi},\tilde{p}_r,\tilde{p}_\phi)}=0\;,\qquad
	\forall x\in\cu{r,\phi,p_r,p_\phi}\;,\quad
	\forall m\in\cu{-1,0,1}\;.
\end{align}
The locus of such points in phase space constitutes the photon shell.
For Schwarzschild, it consists of a single photon sphere.
Since we have specialized to the equatorial plane, we find
\begin{align}
	\tr=3M\;,\qquad
	\tilde{\phi}\in[0,2\pi)\;,\qquad
	\tilde{p}_r=0\;,\qquad
	\tilde{p}_\phi=\pm3\sqrt{3}MH\;,
\end{align}
with the sign corresponding to the prograde/retrograde circular orbit, parameterized by $\tilde{\phi}$.

Importantly, the homoclinic orbits control the long-time behavior of the flow generated by $H_0$.
Under these scalings, $\th$ flows according to $\pd_\alpha\th=\{H_0,\th\}=-\th$, so that after a finite dilation $e^{-\alpha H_0}$,
\begin{align}
	\label{eq:FiniteDilation}
	\th(0)\to\th(\alpha)=e^{-\alpha}\th(0)\;.
\end{align}
For large $\alpha$, $\th$ becomes small, while conversely $T \to\infty$. 
Introducing a dimensionless radius $r=3M\pa{1+R}$, the point of closest approach to the photon shell given in \eqref{eq:RadialTurningPoint} becomes
\begin{align}
	R_{\rm min}^2=-\frac{2\sqrt{3}\th M}{L}+\ldots
\end{align}
to leading order as $\th\to0$. 
It follows that under $\slpr$ dilations,
\begin{align}
	\label{eq:RadialDeviation}
	\pd_\alpha\ln{R_{\rm min}}=-\frac{1}{2}\;.
\end{align}
For $\alpha\to\infty$, $R_{\rm min}\to0$ and $\th\to0$ so any geodesic approaches the bound orbit at $\tr=3M$.
A geodesic that begins and ends at null infinity will do so for any $\alpha$, but since 
\begin{align}
	\label{eq:WindingDivergence}
	\Delta\phi=2\int_{r_{\rm min}}^\infty\frac{L}{r^2\sqrt{\mathcal{V}(r)}}\ed r
	=\log\pa{\frac{1}{R_{\rm min}^2}}+2\log\br{12\pa{2-\sqrt{3}}}+\mathcal{O}\!\pa{R_{\rm min}}
\end{align}
to leading order as $R_{\rm min}\to0$, the number of times $w=\Delta\phi/(2\pi)$ that it orbits the black hole will diverge like\footnote{The variable $w$ interpolates between an integer-spaced set of windings that are inequivalent to the image label $n\in\mathbb{N}$ employed in previous treatments of the photon ring \cite{Johnson2020,Himwich2020,GrallaLupsasca2020a,Hadar2021}.
Rather, the direct ($n=0$) image arises from the unique light ray connecting the source to the observer after executing no more than half a turn around the black hole, i.e., with winding $-\frac{1}{2}\leq w_0\leq\frac{1}{2}$.
Light rays with $n>0$ correspond to relativistic images with
$\Delta w=w-w_0=\sign(w_0)\br{(-1)^n(\frac{n}{2}+\frac{1}{4})-\frac{1}{4}}=\sign(w_0)\cu{-1,1,-2,2,-3,\ldots}$.}
\begin{align}
	\label{eq:WindingExponent}
	\pd_\alpha w=\frac{1}{2\pi}\;.
\end{align}

The emergence of an enhanced $\slr$ symmetry at a fixed point of a scaling transformation is a ubiquitous phenomenon in a wide variety of physical systems.
It is characterized by critical exponents that quantify how various physical quantities scale in the approach to the fixed point.
In the present example, these exponents include the scaling of the radial deviation from the photon shell \eqref{eq:RadialDeviation} or equivalently the orbit number \eqref{eq:WindingExponent}.

Here we are interested in how the critical exponents can be measured astronomically.
In order to discuss this, we introduce a source star at $(r_s,\phi_s)$ and a telescope at $(r_o,\phi_o)$ and consider the (colored) geodesics connecting the two.
There are an infinite number of such geodesics labeled by the number of times $w$ they wind around the black hole en route from star to telescope.
Since each of these geodesics has the same endpoints, they all share the same net angular shift $\Delta\phi$ modulo $2\pi$.

Consider the superselection sector $\Gamma_L=(r,\phi,p_r,p_\phi=L)$ of geodesics with fixed angular momentum $L$.
Demanding that a geodesic in $\Gamma_L$ originate at the star and end at the telescope cuts this three-dimensional subspace of $\Gamma$ down to an infinite but discrete set of geodesics labelled by $w$ and denoted $\Gamma_{\rm obs}$.\footnote{Fixing one endpoint of the geodesic at the star leaves one degree of freedom in $\Gamma_L$: the energy $H$ parameterizing the emission direction via \eqref{eq:SchPr}.
Requiring the geodesic to reach the telescope imposes a condition on $H$ via \eqref{eq:Azimuth}.
Given \eqref{eq:WindingDivergence}, this condition admits a discrete spectrum of solutions labeled by $w$.}
Since $\slpr$ acts transitively on $\Gamma_L$, any two points in $\Gamma_{\rm obs}$ can be related by an $\slpr$ transformation.
However, in general, the set of such transformations do not form a discrete subgroup of $\slpr$.
 
Such a discrete subgroup emerges near the fixed point for $w\gg1$, or equivalently small $R_{\rm min}$.\footnote{A similar argument applies to orbits that wind in the other direction with $w\ll-1$.}
If we act on such a geodesic with a small finite dilation $e^{-\alpha H_0}$, then according to \eqref{eq:WindingExponent}, $\Delta\phi\to\Delta\phi+\alpha$.
It follows that, for the $w$-independent dilation 
\begin{align}
	\label{eq:Dilation}
 	D_0=e^{-2\pi H_0}\;,
\end{align}
we obtain an $\slpr$ element which maps $\Gamma_{\rm obs}$ to itself (for large $w\gg1$) with 
\begin{align}
	w\to w+1\;.
\end{align}
The semigroup formed by products of $D_0$ is an emergent discrete scaling symmetry of the photon ring. 

\begin{figure}[h]
	\centering
	\includegraphics[scale=.5]{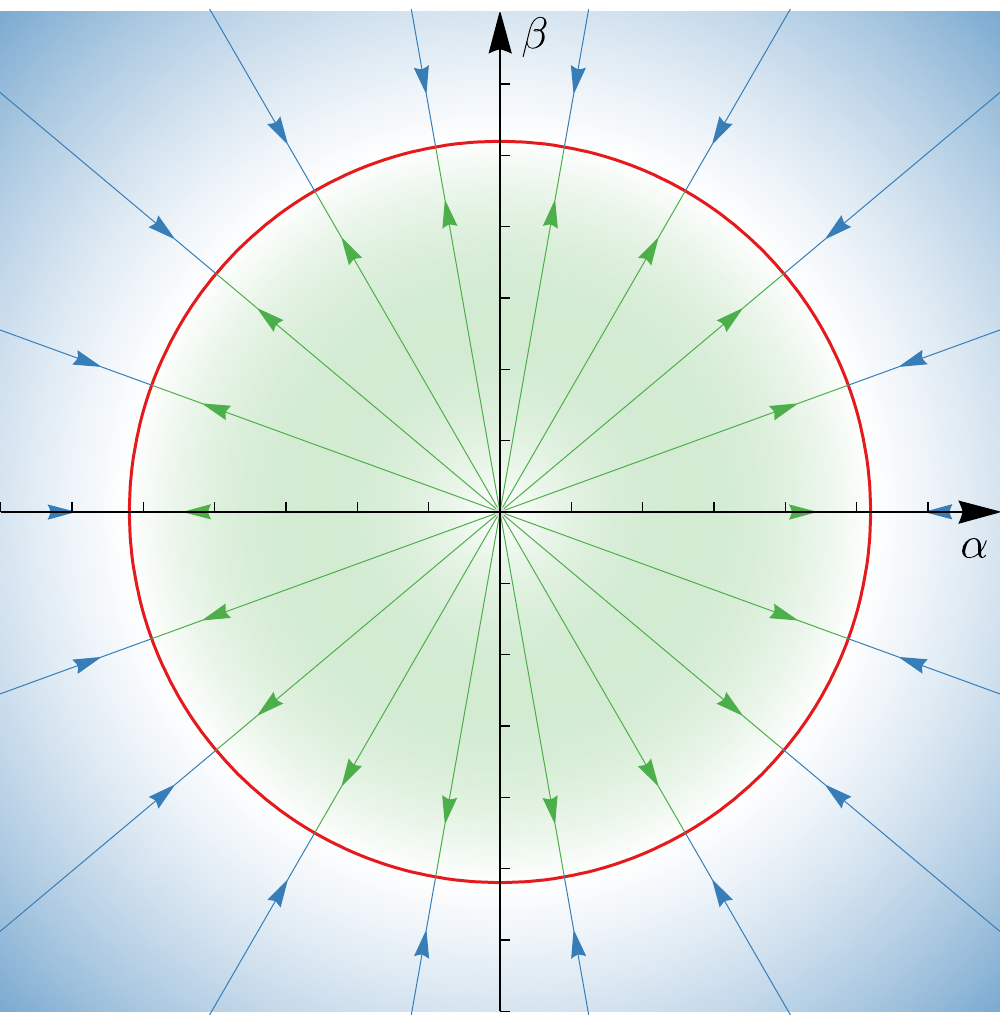}
	\caption{
	Action of $\slpr$ dilations on the image plane of an observer at a large distance from a Schwarzschild black hole.
	}
	\label{fig:Schwarzschild}
\end{figure}

The action of $H_0$ on the observer screen is illustrated in Fig.~\ref{fig:Schwarzschild}.
The red critical curve corresponds to photons with $\th=0$ that asymptote to bound photon orbits at $\tr=3M$.
Photons in the interior of this curve have $\th>0$ and are captured by the black hole, while those in its exterior have $\th<0$ and are deflected back to null infinity.
If the color (energy) of observed photons is fixed, then every choice of a geodesic plane and angular momentum $L$ defines a unique point on the screen (i.e., a direction of approach to the telescope).
However, if $E$ is not fixed, then each choice of a geodesic plane and angular momentum $L$ corresponds to a segment of a ray on the image (either the green or blue segment according to whether $\th\gtrless0$).
Moving along such a segment scales $\th$, keeping the endpoint of the geodesic on the observer screen fixed but varying the other endpoint in the bulk.
Only when $\th$ takes certain discrete values does a photon reach a given source star.
Near the critical curve ($R_{\rm min}\to0$), such photons wind multiple times $\ab{w}\gg1$ around the black hole, and successive images of the star are squeezed exponentially closer to the critical curve by a demagnification factor $e^{-\gamma_L\tau}=e^{-\pi}$, where $\gamma_L$ is the Lyapunov exponent \eqref{eq:SchLyp} and $\tau$ the half-orbital period \eqref{eq:SchOrb}.
These images decompose into two families of photons with different signs of the angular momentum.
Each family consists of a representation of the discrete subgroup of $\slpr$ generated by the finite dilation $D_0$ [Eq.~\eqref{eq:Dilation}].

The preceding discussion did not depend sensitively on the radius of the emitting star and could be any kind of light source outside the photon ring.
In fact, it also applies to light sources appearing inside the photon ring, except that those images arise from photons emitted on geodesics with $\th>0$ that emanate from the past horizon before escaping to future null infinity.
The discrete action of $D_0$ acts on all geodesics that hover near the photon shell and increases their winding number by one.
Hence, acting on the photon ring image at the telescope, it maps the $w^\text{th}$ photon subring to the $(w+1)^\text{th}$ subring.\footnote{Equivalently, in the convention of \cite{Johnson2020}, it maps the $n^\text{th}$ subring to the $(n+2)^\text{th}$ subring.
The full photon ring decomposes into two representations of the discrete subgroup of $\slpr$ generated by $D_0$, corresponding to two sets of lensed images appearing on diametrically opposed positions on the photon ring.}
The radial separation of successive images on the observer screen directly measures the critical exponent $\gamma=\gamma_L\tau$ and therefore the imaginary part of the eikonal quasinormal mode spectrum.
A space-VLBI experiment could therefore directly measure a part of the QNM spectrum that is not accessible by other means, and thereby discover an entirely new emergent symmetry in nature!

\section{Kerr black holes}
\label{sec:Kerr}

We now turn to the astrophysically interesting case of Kerr black holes, which possess a much richer spectrum of bound orbits and quasinormal modes.
The relationship between the Lyapunov exponents of the photon shell and the QNM frequencies in the geometric optics limit was worked out in a series of papers \cite{Goebel1972,Ferrari1984,Mashhoon1985,Iyer1987a,Iyer1987b,Seidel1990,Decanini2003,Dolan2009,Cardoso2009,Dolan2010,Dolan2011,Decanini2010}, culminating in \cite{Yang2012}.
We review and elaborate on this correspondence in sections \ref{subsec:KerrShell} and \ref{subsec:KerrQNM}.
In section \ref{subsec:KerrOvertones}, we describe a novel and very simple geometric optics derivation of the Kerr QNM overtone wavefunctions, and in section \ref{subsec:KerrScreen} we describe the emergent conformal symmetry of the optical image of astrophysical black holes such as M87*. 

\subsection{The near-ring region}
\label{subsec:KerrShell}

The Kerr photon shell is the region of phase space corresponding to (unstably) bound photon orbits that neither escape to infinity nor fall across the event horizon (see, e.g., Fig.~2 of \cite{Johnson2020} for an illustration).
For Schwarzschild, the photon shell can be described in configuration space as a two-sphere at $\tr=3M$.
For a spinning black hole, this sphere radially fattens into a three-dimensional shell.
In Boyer-Lindquist coordinates, each bound orbit has a fixed radial coordinate $r$ but executes complicated motion in the angular directions. 

The Kerr metric for mass $M$ and angular momentum $J=aM$ (with $0\le a\le M$) is
\begin{subequations}
\begin{gather}
	ds^2=-\frac{\Delta}{\Sigma}\pa{\ed t-a\sin^2{\theta}\ed\phi}^2+\frac{\Sigma}{\Delta}\ed r^2+\Sigma\ed\theta^2+\frac{\sin^2{\theta}}{\Sigma}\br{\pa{r^2+a^2}\ed\phi-a\ed t}^2\;,\\
	\Delta=r^2-2Mr+a^2,\qquad
	\Sigma=r^2+a^2\cos^2{\theta}\;.
\end{gather}
\end{subequations}
This geometry has two independent Killing vectors $\pd_\phi$ and $\pd_t$ and an independently conserved  Killing tensor
\begin{align}
	\label{eq:KerrKilling}
	K_{\mu\nu}=-{J_\mu}^\lambda J_{\lambda\nu}\;,\qquad
	J=a\cos{\theta}\ed r\wedge\pa{\ed t-a\sin^2{\theta}\ed\phi}+r\sin{\theta}\ed\theta\wedge\br{\pa{r^2+a^2}\ed\phi-a\ed t}\;.
\end{align}
The corresponding conserved quantities are
\begin{align}
	E=p_\mu\pd_t^\mu
	=-p_t\;,\qquad
	L=p_\mu\pd_\phi^\mu
	=p_\phi\;,\qquad
	k=K^{\mu\nu}p_\mu p_\nu\;.
\end{align}
It is often more convenient to work with the Carter constant $Q=k-\pa{L-aE}^2$.
The local tangent to a null geodesic is completely determined by the conserved quantities $E$, $L$ and $Q$ as 
\begin{align}
    \label{eq:KerrPlow}
	p(x^\mu,E,L,Q)=-E\ed t\pm_r\frac{\sqrt{\mathcal{R}(r)}}{\Delta(r)}\ed r\pm_\theta\sqrt{\Theta(\theta)}\ed\theta+L\ed\phi \;.
\end{align}
The signs $\pm_r$ and $\pm_\theta$ determine the radial and polar directions of travel and the potentials take the form
\begin{subequations}
\begin{align}
    \label{eq:KerrRadialPotential}
	\mathcal{R}(r)&=\br{E\pa{r^2+a^2}-aL}^2-\Delta(r)\br{Q+\pa{L-aE}^2}\;,\\
	\label{eq:KerrPolarPotential}
	\Theta(\theta)&=Q+L^2+a^2E^2\cos^2\theta-\frac{L^2}{\sin^2\theta}\;.
\end{align}
\end{subequations}
Null geodesics with positive $\eta>0$ oscillate in the $\theta$-direction between the zeroes $\theta_\pm$ of the angular potential \eqref{eq:KerrPolarPotential}.

Bound photon orbits require $\mathcal{R}(r)=\mathcal{R}'(r)=0$ and lie at fixed orbital radius $r=\tr$ in the range $\tr\in\br{\tr_-,\tr_+}$ with
\begin{align}
    \tr_\pm=2M\br{1+\cos\pa{\frac{2}{3}\arccos\pa{\pm\frac{a}{M}}}}\;.
\end{align}
Their energy-rescaled angular momentum $\lambda=\frac{L}{E}$ and Carter constant $\eta=\frac{Q}{E^2}$ are fixed by the orbital radius: 
\begin{align}
    \label{eq:CriticalParameters}
	\tilde{\lambda}=a+\frac{\tr}{a}\left[\tr-\frac{2\Delta(\tr)}{\tr-M}\right]\;, \qquad
	\tilde{\eta}=\frac{\tr^3\br{4a^2M-\tr\pa{\tr-3M}^2}}{a^2\pa{\tr-M}^2}\;.
\end{align}
On the boundaries $r=\tr_\pm$, the orbits are equatorial with $\tilde{\eta}=0$, but the bound geodesics in the interior of the photon shell have $\tilde{\eta}>0$ and oscillate in the $\theta$-direction between polar angles given by
\begin{align}
    \label{eq:CriticalPoles}
    \tilde{\theta}_\pm=\arccos\pa{\mp\sqrt{\tilde{u}_+}}\;,\qquad
    \tilde{u}_\pm=\frac{-\tr^4+3M^2\tr^2-2a^2M\tr\pm2\tr\sqrt{M\Delta(\tr)\pa{2\tr^3-3M\tr^2+a^2M}}}{a^2\pa{\tr-M}^2}
    \gtrless0\;.
\end{align}
Following \cite{Johnson2020}, we will refer to one such complete oscillation (e.g., from $\tilde{\theta}_-$ back to itself) as one orbit, since the photon typically returns to a point near, but not identical to (since the angle $\phi$ also shifts), its initial position. Note that $\tilde{\theta}_\pm$ tends to the north/south poles (i.e., the bound orbit passes over the poles) if and only if $\tilde{\lambda}=0$.
The angular momentum vanishes at the orbital radius $\tr_0\in\br{\tr_-,\tr_+}$ given by
\begin{align}
    \tr_0=M+2M\triangle\cos\br{\frac{1}{3}\arccos\pa{\frac{1-\frac{a^2}{M^2}}{\triangle^3}}}\;,\qquad
    \triangle=\sqrt{1-\frac{a^2}{3M^2}}\;.
\end{align}

The bound geodesics are unstable in the sense that any small perturbation will push them into the black hole or towards infinity where they can reach a telescope.
The observed photon shell image arises from photons traveling on such ``nearly bound'' geodesics.
Consider two such nearby geodesics, one of which is exactly bound at $\tr$, with the other initially differing only by an infinitesimal radial separation $\dt r_0$.
After $n$ half-orbits the separation grows as
\begin{align}
	\dt r_n=e^{\gamma n}\dt r_0\;,
\end{align} 
where the Lyapunov exponent $\gamma$ is a function on the space of bound orbits given by 
\begin{align}
	\label{eq:KerrLyp}
	\gamma=\sqrt{\frac{\mathcal{R}''(\tr)}{2E^2}}\tilde{G}_\theta\;,\qquad
	\tilde{G}_\theta=\int_{\tilde{\theta}_-}^{\tilde{\theta}_+}\frac{E}{\sqrt{\Theta(\theta)}}\ed\theta
	=\frac{2}{\sqrt{-\tilde{u}_-a^2}}K\!\pa{\frac{\tilde{u}_+}{\tilde{u}_-}}\;.
\end{align}
The elliptic integral arises from averaging over an angular period.
The half-period of a bound orbit at $r=\tr$ is \cite{GrallaLupsasca2020a}
\begin{align}
    \tau=\tr^2\pa{\frac{\tr+3M}{\tr-M}}\tilde{G}_\theta+a^2\tilde{G}_t\;,\qquad
    \tilde{G}_t=\int_{\tilde{\theta}_-}^{\tilde{\theta}_+}\frac{E\cos^2{\theta}}{\sqrt{\Theta(\theta)}}\ed\theta
    =2\sqrt{-\frac{\tilde{u}_-}{a^2}}\br{E\!\pa{\frac{\tilde{u}_+}{\tilde{u}_-}}-K\!\pa{\frac{\tilde{u}_+}{\tilde{u}_-}}}\;.
\end{align}
The period-averaged radial deviation as a function of Boyer-Lindquist time becomes \cite{Yang2012}
\begin{align}
    \label{eq:KerrRadialDeviation}
	\dt r(t)=e^{\gamma_Lt}\dt r_0\;,\qquad	\gamma_L=\frac{\gamma}{\tau}\;.
\end{align}

\subsection{Conformal symmetry of the quasinormal mode spectrum}
\label{subsec:KerrQNM}

We now turn to a discussion of the quasinormal modes of the Kerr black hole.
The wave equation \eqref{eq:Wav} in this geometry separates due to the existence of the Killing tensor \eqref{eq:KerrKilling}.
Solutions to the massless scalar wave equation therefore take the form 
\begin{align}\label{eq:KerrQNMansatz}
    \Phi(t,r,\theta,\phi)=\int\ed\omega\sum_{\ell=0}^\infty\sum_{m=-\ell}^{\ell}c_{\ell m}(\omega)\Phi_{ \ell m \omega}(t,r,\theta,\phi)\;, \qquad 
    \Phi_{\ell m\omega}(t,r,\theta,\phi)=e^{-i\omega t}R_{\ell m\omega}(r)S_{\ell m\omega}(\theta)e^{im\phi}\;,
\end{align}
where the $S_{\ell m\omega}(\theta)$ denote the spheroidal harmonics.
These special functions satisfy the equation 
\begin{align}
    \label{eq:KerrAngularEq}
    \br{\frac{1}{\sin{\theta}}\pd_\theta\pa{\sin{\theta}\pd_\theta}+a^2\omega^2\cos^2\theta -\frac{m^2}{\sin^2\theta}+A_{\ell m}}S_{\ell m\omega}(\theta)=0\;,
\end{align}
where the $A_{\ell m}(a\omega)$ are angular separation constants, which are quantized by requiring $S_{\ell m\omega}(\theta)$ to be regular at the poles.
Note that the potential in this ODE matches the geodesic polar potential \eqref{eq:KerrPolarPotential} provided that we identify
\begin{align}
    \label{eq:GeoIdf}
    \omega=E\;,\qquad
    m=L\;,\qquad
    A_{\ell m}=Q+m^2\;.
\end{align}
The $S_{\ell m\omega}(\theta)$ are labelled by the integers $\ell$ and $m$ with $-\ell\le m\le\ell$ and depend explicitly on the quantity $a^2\omega^2$.
For a round sphere (as in Schwarzschild), the separation constants  are $A_{\ell m}=\ell(\ell+1)$, but there is no analytic formula for the $A_{\ell m}$ in generic Kerr and they must be computed numerically.%\footnote{In \texttt{Mathematica}, the spheroidal harmonics are implemented as $S_{\ell m\omega}(\theta)=\mathsf{SpheroidalPS}[\ell,m,ia\omega,\cos{\theta}]$ and the separation constants as $A_{\ell m}+(a\omega)^2=\mathsf{SpheroidalEigenvalue}[\ell,m,ia\omega]$.}
The constants $A_{\ell m}=A_{\ell m}^R+iA_{\ell m}^I$ are in general complex (as is $\omega=\omega_R+i\omega_I$), but in the eikonal approximation the real parts dominate the imaginary parts.
Since the angular separation constants enter the radial equation, a separate WKB analysis must also be performed on \eqref{eq:KerrAngularEq}.
Defining $e^y=\tan{\frac{\theta}{2}}$, so that $\pd_y=\sin{\theta}\pd_\theta$, the equation becomes
\begin{align}
    \br{\frac{1}{\sin^2{\theta}}\pd_y^2+a^2\omega_R^2\cos^2\theta -\frac{m^2}{\sin^2\theta}+A^R_{\ell m}}S_{\ell m\omega}(y)=-i\pa{2\omega_R\omega_Ia^2\cos^2{\theta}+A^I_{\ell m}}S_{\ell m\omega}(y)\;.
\end{align}
 We assume that $A^R_{\ell m}\gg A^I_{\ell m}$ as well as $\omega_R\gg\omega_I$, with $A_{\ell m}^R\sim \omega_R^2\sim m^2\sim\pa{A_{\ell m}^I}^2$ of comparable magnitudes. In this regime, a WKB analysis of the equation then yields \cite{Yang2012,Yang2014}  
\begin{align}
    \label{eq:SepConst}
    A^I_{\ell m}=-2a^2\omega_R\omega_I\frac{\tilde{G}_t}{\tilde{G}_\theta}\;.
\end{align}
The radial part of the scalar wave equation takes the form
\begin{align}
    \label{eq:KerrRadialOperator}
    \br{\Delta(r)\pd_r\pa{\Delta(r)\pd_r}+V(r)}R_{\ell m\omega}(r)=0\;,
\end{align}
where the wave potential is given by
\begin{align}
    V(r)=\br{\omega\pa{r^2+a^2}-am}^2-\Delta(r)\pa{A_{\ell m}+a^2\omega^2-2am\omega}\;.
\end{align}
Again, note that the potential in this ODE matches the geodesic radial potential \eqref{eq:KerrRadialPotential} provided that we make the same identifications as in \eqref{eq:GeoIdf}.
In terms of the tortoise coordinate defined by $\frac{dr_*}{dr}=\frac{r^2+a^2}{\Delta(r)}$ and the rescaled radial function $\psi_{\ell m\omega}(r)=\sqrt{r^2+a^2}R_{\ell m\omega}(r)$, this radial ODE becomes
\begin{align}
    \label{eq:KerrExact}
    \br{\pd_{r_*}^2+\frac{V(r)}{\pa{r^2+a^2}^2}-\frac{g(r)\Delta(r)}{\pa{r^2+a^2}^4}}\psi_{\ell m\omega}(r_*)=0\;,\qquad
    g(r)=2Mr^3+a^2r(r-4M)+a^4 \;.
\end{align}
We are interested in the geometric optics limit, where analytic formulas are available.
This limit is $\ell\to\infty$ with
\begin{align}
	\label{eq:KerrEikonal}
	\mu\equiv\frac{m}{\ell+\frac{1}{2}}\in\pa{-1,1}\;,\qquad
	\Omega_R(\mu)\equiv\frac{\omega_R}{\ell+\frac12}\;,\qquad
	A(\mu)\equiv\frac{A_{\ell m}^R}{\pa{\ell+\frac{1}{2}}^2}\;,\qquad
	\frac{A_{\ell m}^I}{\pa{\ell+\frac{1}{2}}}\;,
\end{align}
all held fixed. As demonstrated in \cite{Yang2012}, at leading order in this limit, a Kerr quasinormal mode with real energy $\omega_R$, angular momentum $m$, and real separation constant $A_{\ell m}^R$ can be associated with a homoclinic congruence of null geodesics with energy $E=\omega_R$, angular momentum $L=m$, and Carter constant $Q=A_{\ell m}^R-m^2$, that asymptote to bound orbits at photon shell radius $\tr$ such that\footnote{This is an excellent approximation to the exact bijection between $\mu$ and $\tr$ obtained from the Bohr-Sommerfeld quantization condition $\int_{\tilde{\theta}_-}^{\tilde{\theta}_+}\sqrt{a^2\omega^2\cos^2{\theta}-\frac{m^2}{\sin^2{\theta}}+A_{\ell m}}\ed\theta=\pi\pa{\ell+\frac{1}{2}-\ab{m}}$ \cite{Yang2012}, which in the eikonal limit implies that $\mu(\tr)=\pa{\sign(\tr_0-\tr)+\frac{\tilde{J}_\theta}{\pi\tilde{\lambda}}}^{-1}$.
Here $\tilde{J}_\theta\equiv\frac{1}{E}\int_{\tilde{\theta}_-}^{\tilde{\theta}_+}\sqrt{\Theta(\theta)}\ed\theta$ is the action integral for the polar motion, which can be evaluated in terms of elliptic integrals as \cite{Kapec2020}
\begin{align*}
    \tilde{J}_\theta=\int_{\tilde{\theta}_-}^{\tilde{\theta}_+}\sqrt{\tilde{\eta}^2+\tilde{\lambda}^2+a^2\cos^2{\theta}-\frac{\tilde{\lambda}^2}{\sin^2{\theta}}}\ed\theta
    =\frac{2}{\sqrt{-\tilde{u}_-a^2}}\br{a^2\pa{1-\tilde{u}_+}K\!\pa{\frac{\tilde{u}_+}{\tilde{u}_-}}-a^2\tilde{u}_-E\!\pa{\frac{\tilde{u}_+}{\tilde{u}_-}}-\tilde{\lambda}^2\Pi\!\pa{\tilde{u}_+;\frac{\tilde{u}_+}{\tilde{u}_-}}}.
\end{align*}
}
\begin{align}
    \label{eq:Bijection}
    \mu(\tr)=\sign(\tr_0-\tr)\sin{\tilde{\theta}_\pm}
    =\sign(\tr_0-\tr)\sqrt{1-\tilde{u}_+(\tr)}
    \in\br{-1,1}\;.
\end{align}
This correspondence determines the QNM wavefronts, while the subleading equation \eqref{eq:GeoAmp} determines their amplitude in terms of the expansion of the null congruence.
The amplitude decay (overtone structure) is determined by expanding \eqref{eq:KerrExact}, again assuming that $A^R_{\ell m}\gg A^I_{\ell m}$ and $\omega_R\gg\omega_I$ with $A_{\ell m}^R\sim \omega_R^2\sim m^2\sim\pa{A_{\ell m}^I}^2$ of comparable magnitudes:
\begin{align}
    &\br{\pd_{r_*}^2+\frac{\br{\omega_R\pa{r^2+a^2}-am}^2-\Delta(r)\pa{A_{\ell m}^R+a^2\omega_R^2-2am\omega_R}}{(r^2+a^2)^2}}\psi_{\ell m\omega}(r_*)\notag\\
    &\qquad\qquad=-i\br{2\omega_R\omega_I-\frac{2am\omega_I}{r^2+a^2}-\frac{\Delta(r)\pa{A_{\ell m}^I+2a^2\omega_R\omega_I-2am\omega_I}}{\pa{r^2+a^2}^2}}\psi_{\ell m\omega}(r_*)\;.
\end{align}
This is still an intractable ODE.
However, it simplifies dramatically in the Kerr near-ring region defined for each orbital radius $\tr\in\br{\tr_-,\tr_+}$ and radial deviation $\dt r=r-\tr$ by
\begin{align}
    \label{eq:NearRingRegionKerrQNMs}
    \text{NEAR-RING REGION:}\qquad
    \begin{cases}
        \ab{\dt r}\ll M&\qquad\text{(near-peak)},\\
        \displaystyle\ab{\frac{m}{\omega_R}-\tilde{\lambda}}\ll M&\qquad\text{(near-critical in $m$ and $\phi$)},\vspace{3pt}\\
        \displaystyle\ab{\frac{A_{\ell m}^R-m^2}{\omega_R^2}-\tilde{\eta}}\ll M&\qquad\text{(near-critical in $\ell$ and $\theta$)},\vspace{3pt}\\
        \displaystyle\frac{1}{\omega_R}\ll M&\qquad\text{(high-frequency)},
    \end{cases}
\end{align}
where at leading order, we can approximate the LHS by a quadratic potential in $\dt r$ while treating the radius on the RHS as a constant $r=\tr$.
In this way, we recover an inverted harmonic oscillator eigenvalue problem as in section ~\ref{sec:Schwarzschild}. Using \eqref{eq:SepConst}, this becomes 
\begin{align}
    \frac{1}{2\omega_R}\br{\pd_{r_*}^2+\frac{1}{2}\frac{\mathcal{R}''(\tr)}{\pa{\tr^2+a^2}^2}\dt r^2}\psi_{\ell m\omega}(r_*)
    &=-i\omega_I\br{1-\frac{a\tilde{\lambda}}{\tr^2+a^2}+\frac{a\Delta(\tr)}{\pa{\tr^2+a^2}^2}\pa{\tilde{\lambda}-a+a\frac{\tilde{G}_t}{\tilde{G}_\theta}}}\psi_{\ell m\omega}(r_*)\\
    &=-i\omega_I\br{\frac{\Delta(\tr)}{\pa{\tr^2+a^2}^2}\frac{\tau}{\tilde{G}_\theta}}\psi_{\ell m\omega}(r_*)\;.
\end{align}
Noting that near the ring $\pd_{r_*}=\frac{\Delta(\tr)}{\tr^2+a^2}\pd_r$, this can be rewritten as 
\begin{align}
    \mathcal{H}\psi=i\omega_I\psi\;,\qquad
    \mathcal{H}=-\frac{\tilde{G}_\theta\pa{\tr^2+a^2}^2}{2\omega_R\tau\Delta(\tr)}\br{\pd_x^2+\pa{\frac{\gamma}{\tilde{G}_\theta}\frac{\Delta(\tr)}{\pa{\tr^2+a^2}^2}}^2\omega_R^2x^2}\;,\qquad
    x=r_*-\tr_*
    =\frac{\tr^2+a^2}{\Delta(\tr)}\dt r\;.
\end{align}
If we impose quasinormal mode boundary conditions, then we find that the eigenvalues of $\mathcal{H}$ must be imaginary with
\begin{align}
    \omega_I=-\pa{n+\frac{1}{2}}\frac{\gamma}{\tau}\;,
\end{align}
in complete agreement with the overtone structure expected from \eqref{eq:KerrRadialDeviation}.
Note that in Kerr, the specific normalization of the Hamiltonian (which is determined by the eikonal form of the angular separation constants) accounts for the $\theta$-averaged motion that defines the Lyapunov exponent.
We next define $k=\frac{\Delta(\tilde{r})}{\tilde{G}_\theta\pa{\tr^2+a^2}^2}$ and the operators 
\begin{align}
    \label{eq:KerrGenerators}
    a_\pm=\frac{e^{\pm\gamma_Lt}}{\sqrt{2k\gamma\omega_R}}\pa{\mp i\pd_x-k\gamma\omega_Rx}\;,\qquad
    L_0&=-\frac{i}{4}\pa{a_+a_-+a_-a_+}
	=\frac{i}{2\gamma_L}\mathcal{H}\;,\qquad
	L_\pm=\pm\frac{a_\pm^2}{2}\;.
\end{align}
As in Schwarzschild, the $a_\pm$ obey the Heisenberg algebra $\br{a_+,a_-}=iI$, while the $L_m$ obey the exact $\slqn$ commutation relations.
Of course, the $L_m$ only commute with the wave equation in the near-ring region, in a particular superselection sector labelled by $(\omega_R,\ell,m)$.
Eigenstates of $L_0$ satisfy $L_0\psi_h=h\psi_h$, so we identify $\omega_I=-2\gamma_L h$.
In the near-ring region, the mode ansatz \eqref{eq:KerrQNMansatz} reduces to
\begin{align}
    \Phi_{\ell m\omega}(t,r,\theta,\phi)\approx e^{-i\omega_Rt}\frac{\Phi_h(t,x)}{\sqrt{r^2+a^2}}S_{\ell m\omega}(\theta)e^{im\phi}\;,\qquad
    \Phi_h(t,x)=e^{\omega_It}\psi_h(x)
    =e^{-2\gamma_L ht}\psi_h(x)\;.
\end{align}
In this framework, the quasinormal mode boundary condition is equivalent to the imposition of a highest-weight condition $L_+\psi_h=0$ on the fundamental mode.
There are two solutions with $h=\frac{1}{4}$ and $h=\frac{3}{4}$:
\begin{align}
    \label{eq:KerrHighestWeight}
    \Phi_\frac{1}{4}(t,x)&=e^{-\frac{1}{2}\gamma_Lt}\psi_\frac{1}{4}(x)\;,
    &&\psi_\frac{1}{4}(x)=e^{\frac{i}{2}k\gamma\omega_Rx^2}\;,\\
    \Phi_\frac{3}{4}(t,x)&=e^{-\pa{1+\frac{1}{2}}\gamma_L t}\psi_\frac{3}{4}(x)\;,
    &&\psi_\frac{3}{4}(x)=xe^{\frac{i}{2}k\gamma\omega_Rx^2}\;.
\end{align}
Higher overtones are then obtained as $\slqn$-descendants: 
\begin{align}
    \Phi_{h,N}(t,x)=L_-^N\Phi_h(t,x)
    =e^{-2\gamma_L(h+N)t}\psi_{h+N}(x)
    \propto e^{-2\gamma_L(h+N)t}D_{2(h+N)-\frac{1}{2}}\pa{\sqrt{-2ik\gamma\omega_R}x}\;,
\end{align}
where $D_n(x)$ denotes the $n^\text{th}$ parabolic cylinder function.
Thus, the QNM overtones in the Kerr near-ring region \eqref{eq:NearRingRegionKerrQNMs} fall into two irreps of the $\slr$ generated by \eqref{eq:KerrGenerators}, obtained from the primary states with $h=\frac{1}{4}$ and $h=\frac{3}{4}$.
The Casimir is 
\begin{align}
    \mathcal{C}\Phi_{h,N}\equiv\pa{-L_0^2+\frac{L_+L_-+L_-L_+}{2}}\Phi_{h,N}=h(1-h)\Phi_{h,N}\;,
\end{align}
so, as in Schwarzschild, the two representations that appear are shadows of each other with Casimir
\begin{align}
    \mathcal{C}=\frac{3}{16}\;.
\end{align}
Finally, we note that at the edges $x\to\pm\infty$ of the near-peak region, 
\begin{align}
    D_n\pa{\sqrt{-2ik\gamma\omega_R}x}\stackrel{x\to\pm\infty}{\sim}x^ne^{\frac{i}{2}k\gamma\omega_Rx^2}\;.
\end{align}
Therefore, the $n^\text{th}$ overtone behaves as 
\begin{align}
    \label{eq:KerrOvt}
    \Phi_{\ell mn}(t,r,\theta,\phi)\stackrel{x\to\pm\infty}{\sim}
    e^{-\pa{n+\frac{1}{2}}\gamma_Lt}x^ne^{-i\omega_R\pa{t-\frac{1}{2}k\gamma x^2}}S_{\ell m\omega}(\theta)e^{im\phi}\;,\qquad
    \omega_R=\pa{\ell+\frac{1}{2}}\Omega_R(\mu)\;.
\end{align}

To summarize, the QNM spectrum $\omega_{\ell mn}$ is labelled by the overtone number $n\in\mathbb{N}$, which is quantized by the quasinormal boundary condition, as well as by the spheroidal number $\ell$ and the angular frequency $m$.
The eikonal limit of the spectrum takes the form 
\begin{align}
    \label{eq:KerrSpectrum}
	\omega_{\ell\mu n}\stackrel{\ell\to\infty}{\approx}\pa{\ell+\frac{1}{2}}\Omega_R(\mu)-i\pa{n+\frac{1}{2}}\gamma_L(\mu)\;,
\end{align}
and the geometric optics approximation determines $\Omega_R(\mu)$ and $\gamma_L(\mu)$ in terms of the geometric data of the photon shell: inverting the bijective function \eqref{eq:Bijection} for $\tr(\mu)$, we have
\begin{align}
	\Omega_R(\mu)=\frac{\mu}{\tilde{\lambda}(\tr(\mu))}\;,\qquad
	\gamma_L(\mu)=\frac{\gamma(\tr(\mu))}{\tau(\tr(\mu))}\;.
\end{align}
The higher overtones are $\slqn$-descendants and the fundamental mode is highest-weight in the near-ring region.
It is a challenge for any proposed holographic dual to Kerr to reproduce the dispersion relation \eqref{eq:KerrSpectrum}.

\subsection{Quasinormal modes from geometric optics}
\label{subsec:KerrOvertones}

Next, we turn to a brief discussion of the QNM wavefunctions and their description within the geometric optics approximation.
The Hamilton-Jacobi principal function for null geodesics in Kerr is
\begin{align}
    \label{eq:HamiltonJacobiKerr}
    S(t,r_*,\theta,\phi)=-Et+L\phi+\int^{r_*}\frac{\sqrt{\mathcal{R}(r_*')}}{\pa{r_*'^2+a^2}}\ed r_*'+\int^\theta\sqrt{\Theta(\theta')}\ed\theta'\;,
\end{align}
which reproduces \eqref{eq:KerrPlow} via $p_\mu=\pd_\mu S(x)$.
As is apparent from \eqref{eq:HamiltonJacobiKerr}, the angular modes $S_{\ell m\omega}(\theta)$ of the QNM wavefunctions are oscillatory in the region between the angular turning points of the corresponding bound null geodesic at $r=\tr(\mu)$ and decay exponentially toward the poles.
This behavior corresponds to the angular trapping of photon shell geodesics, which are confined to oscillate about the equatorial plane.
In the near-ring region, the radial potential in \eqref{eq:HamiltonJacobiKerr} exhibits a double zero and integrates to the phase in \eqref{eq:KerrHighestWeight}.
The leading ($\theta$-averaged) solution to \eqref{eq:GeoAmp} is simply $A_0=e^{-\frac{1}{2}\gamma_Lt}$, which when combined with the critical phase $\tilde{S}$ in \eqref{eq:HamiltonJacobiKerr}, gives an excellent approximation to the fundamental quasinormal mode. 

As discussed in section \ref{sec:QNMGeo}, it is possible to construct additional solutions to the geometric optics wave equation given the fundamental seed solution $\Phi_0(x)=A_0(x)e^{i\tilde{S}(x)}$ and a solution to the homogeneous amplitude equation
\begin{align}
    p^\mu\pd_\mu u=0\;,
\end{align}
in this case averaged over the polar motion. 
According to \eqref{eq:KerrRadialDeviation}, the quantity $u=e^{-\gamma_Lt}(r-\tr)$ is conserved along the homoclinic trajectories after an appropriate averaging over the polar motion.
We can therefore write down a family of solutions associated to the same critical orbit with phase $\tilde{S}$, but differing in the amplitude
\begin{align}
    \Phi_n(x)\sim e^{-\pa{n+\frac{1}{2}}\gamma_Lt}\pa{r-\tr}^ne^{i\tilde{S}(x)}\;.
\end{align}
This eikonal QNM approximation agrees with the near-ring approximation \eqref{eq:KerrOvt} in their overlap region. 

\subsection{Observable conformal symmetry of the photon ring}
\label{subsec:KerrScreen}

In this subsection, we identify the emergent near-ring conformal symmetry $\slpr$ for the Kerr black hole.
This structure is far richer in Kerr than in Schwarzschild, and a measurement of its critical exponents would provide a sensitive probe of spinning black holes.
The intricacy of the Kerr lens will lead us to consider stationary, axisymmetric, fixed-$\theta$ source rings rather than pointlike sources, in order to allow a simplified study of the consequences of the conformal symmetry on black hole images.
For time-averaged images, this restriction is well-motivated observationally.

Let $\Gamma$ denote the six-dimensional phase space of colored null geodesics in Kerr, spanned by $(r,\theta,\phi,p_r,p_\theta,p_\phi)$ with canonical symplectic form.
Time evolution is generated by the Hamiltonian 
\begin{gather}
	\label{eq:KerrH}
	H(r,\theta,p_r,p_\theta,p_\phi)=\br{\frac{\pa{r^2+a^2}^2}{\Delta(r)}-a^2\sin^2{\theta}}^{-1}\pa{\frac{2Mar}{\Delta(r)}p_\phi+\sqrt{G}}\; ,\\
	G=\pa{\frac{2Mar}{\Delta(r)}p_\phi}^2+\br{\frac{\pa{r^2+a^2}^2}{\Delta(r)}-a^2\sin^2{\theta}}\br{\Delta(r)p_r^2+p_\theta^2+\pa{\frac{1}{\sin^2{\theta}}-\frac{a^2}{\Delta(r)}}p_\phi^2}\;,
\end{gather}
which is obtained by solving the null condition $g^{\mu\nu}p_\mu p_\nu=0$ for $p_t=-H$.
The Carter constant
\begin{align}
	\label{eq:RadialCarter}
	Q(r,\theta,p_r,p_\theta,p_\phi)
	&=-\Delta(r)p_r^2+\frac{\br{H\pa{r^2+a^2}-ap_\phi}^2}{\Delta(r)}-\pa{p_\phi-aH}^2\\
	\label{eq:AngularCarter}
	&=p_\theta^2-a^2H^2\cos^2{\theta}+p_\phi^2\cot^2{\theta}
\end{align}
commutes with the Hamiltonian \eqref{eq:KerrH} and is therefore conserved along each photon trajectory, as is the angular momentum $L=p_\phi$.
Inverting \eqref{eq:RadialCarter} and \eqref{eq:AngularCarter} respectively gives
\begin{align}
	\label{eq:RadialMomentum}
	p_r(r,H,L,Q)&=\frac{\pm\sqrt{\mathcal{R}(r)}}{\Delta(r)}\;,&&
	\mathcal{R}(r)=\br{H\pa{r^2+a^2}-aL}^2-\Delta(r)\br{Q+\pa{L-aH}^2}\;,\\
	\label{eq:PolarMomentum}
	p_\theta(\theta,H,L,Q)&=\pm\sqrt{\Theta(\theta)}\;,&&
	\Theta(\theta)=Q+L^2+a^2H^2\cos^2{\theta}-\frac{L^2}{\sin^2{\theta}}\;.
\end{align}
The coordinate transformation $(r,\theta,\phi,p_r,p_\theta,p_\phi)\to\pa{T,\Phi,\Psi,H,L,Q+L^2}$ defined by 
\begin{align}
	\ed T&=\frac{Hr^2\Delta+2Mr\br{H\pa{r^2+a^2}-aL}}{\Delta(r)\sqrt{\mathcal{R}(r)}}\ed r+\frac{a^2H\cos^2{\theta}}{\sqrt{\Theta(\theta)}}\ed\theta\;,\\
	\ed\Phi&=\ed\phi-\br{\frac{a\pa{2HMr-aL}}{\Delta(r)\sqrt{\mathcal{R}(r)}}\ed r+\frac{L\csc^2{\theta}}{\sqrt{\Theta(\theta)}}\ed\theta}\;,\\
	\ed\Psi&=-\frac{1}{2}\br{\frac{1}{\sqrt{\mathcal{R}(r)}}\ed r-\frac{1}{\sqrt{\Theta(\theta)}}\ed\theta}\;,
\end{align}
is canonical since it preserves the symplectic form
\begin{align}
	\Omega=\ed p_r\wedge\ed r+\ed p_\theta\wedge\ed\theta+\ed p_\phi\wedge\phi
	=\ed H\wedge\ed T+\ed L\wedge\ed\Phi+\ed\pa{Q+L^2}\wedge\ed\Psi\;.
\end{align}
These canonical coordinates lead to trivial equations of motion for the Hamiltonian \eqref{eq:KerrH}:
\begin{align}
	\dot{H}&=\cu{H,H}=0\;,
	&&\dot{L}=\cu{L,H}=0\;,
	&&\dot{Q}=\cu{Q,H}=0\;,\\
	\dot{\Psi}&=\cu{\Psi,H}=0\;,
	&&\dot{\Phi}=\cu{\Phi,H}=0\;,
	&&\dot{T}=\cu{T,H}=1\;.
\end{align}
The first three equations indicate that the phase space $\Gamma$ foliates into superselection sectors of fixed $(H,L,Q)$, which are conserved momenta.
The fourth and fifth equations imply that 
the Hamiltonian flow sends a photon with initial coordinates $(r_s,\theta_s,\phi_s,H,L,Q)$ to final coordinates $(r_o,\theta_o,\phi_o,H,L,Q)$ according to the rule
\begin{gather}
	\label{eq:MinoTime}
	\fint_{r_s}^{r_o}\frac{\ed r}{\sqrt{\mathcal{R}(r)}}=\fint_{\theta_s}^{\theta_o}\frac{\ed\theta}{\sqrt{\Theta(\theta)}}\;,\\
	\label{eq:KerrAzimuth}
	\Delta\phi=\phi_o-\phi_s
	=\fint_{\phi_s}^{\phi_o}\ed\phi
	=\fint_{r_s}^{r_o}\frac{a\pa{2HMr-aL}}{\Delta(r)\sqrt{\mathcal{R}(r)}}\ed r+\fint_{\theta_s}^{\theta_o}\frac{L\csc^2{\theta}}{\sqrt{\Theta(\theta)}}\ed\theta\;,
\end{gather}
where the slash indicates that an integral is to be evaluated along the photon trajectory.
Finally, the last equation identifies $T$ as the variable conjugate to energy, i.e., time.
Hence, the time elapsed during evolution from a state $(r_s,\theta_s,\phi_s,H,L,Q)$ to $(r_o,\theta_o,\phi_o,H,L,Q)$ is
\begin{align}
	\label{eq:KerrTimeLapse}
	T&=\fint_{r_s}^{r_o}\frac{Hr^2\Delta+2Mr\br{H\pa{r^2+a^2}-aL}}{\Delta(r)\sqrt{\mathcal{R}(r)}}\ed r+\fint_{\theta_s}^{\theta_o}\frac{a^2H\cos^2{\theta}}{\sqrt{\Theta(\theta)}}\ed\theta\;.
\end{align}
Equations \eqref{eq:MinoTime}, \eqref{eq:KerrAzimuth} and \eqref{eq:KerrTimeLapse} are the solution to the null geodesic equation in Kerr.
These integrals can be evaluated explicitly in terms of elliptic functions \cite{GrallaLupsasca2020b}, but their detailed form will not be needed here.
As in Schwarzschild, the salient feature of \eqref{eq:KerrTimeLapse} is the logarithmic divergence along the homoclinic trajectories associated to the double zero of the radial potential.
This $T$ is a local coordinate, and it becomes singular in the vicinity of a hyperbolic fixed point. 

Bound photon orbits occur in the range $\tr_-\leq\tr\leq\tr_+$, in which $\dot{r}=\dot{p}_r=0$ can vanish simultaneously.
The conserved quantities associated to these orbits are determined by the conditions $\mathcal{R}(\tr)=\mathcal{R}'(\tr)=0$ that define the photon shell in phase space.
The energy-rescaled critical parameters $(\tilde{\lambda},\tilde{\eta})$ are given by the relations \eqref{eq:CriticalParameters}, which can be inverted to obtain $\tr(L,Q)$, and thence the zero-point energy $\tilde{H}(L,Q)$.

As in the Schwarzschild case, we now define $\th=H-\tilde{H}$ and consider the unbound geodesics with $\th<0$ that begin and end at null infinity, always remaining outside the black hole.
A distant observer at large radius $r_o\to\infty$ receives such a geodesic with impact parameters $(\lambda,\eta)$ at the position $(\alpha,\beta)$ on the observer sky given by
\begin{align}
    \label{eq:BardeenCoordinates}
    \alpha=-\frac{\lambda}{\sin{\theta_o}}\;,\qquad
    \beta=\pm\sqrt{\eta+a^2\cos^2{\theta_o}-\lambda^2\cot^2{\theta_o}}\;.
\end{align}
For such a (non-homoclinic) geodesic, the radius of closest approach to the black hole is attained when the radial momentum \eqref{eq:RadialMomentum} vanishes.
This occurs at the largest real root of the quartic potential $\mathcal{R}(r)$, which is given explicitly in Eq.~(95d) of \cite{GrallaLupsasca2020b}.
Geodesics with $\th=0$ asymptote to bound photon orbits at $r=\tr(L,Q)$ in the far past or future.
Their impact parameters $(\tilde{\alpha},\tilde{\beta})$, obtained by substituting \eqref{eq:CriticalParameters} into \eqref{eq:BardeenCoordinates}, define the Kerr critical curve $\mathcal{C}(\tr)$ in the observer sky. 

Since the coordinates $\pa{T,\Phi,\Psi,H,L,Q+L^2}$ are canonical,  the functions\footnote{One can generalize $\th$ to any function of the form $H-g(L,Q)$ for some $g$, but only the choice in \eqref{eq:GeneratorsKerr} leads to dilations that scale onto homoclinic orbits.
One could also add a Casimir $\frac{\mathcal{C}(L,Q)}{\th}$ to $H_-$, but we will not use special conformal transformations explicitly here.}
\begin{align}
	\label{eq:GeneratorsKerr}
	H_+=\th\;,\qquad
	H_0=-\th T\;,\qquad
	H_-=\th T^2\;,
\end{align}
obey the $\slpr$ algebra.
This $\slpr$ commutes with both $L$ and $Q$ and therefore acts within superselection sectors $\Gamma_{L,Q}$ of fixed angular momentum and Carter constant.
However, the flow generated by $H_0$ does modify the energy (or photon color) $H=\th+\tilde{H}(L,Q)$ and therefore
acts on the impact parameters as well as the radius of closest approach.
The action on $\Gamma_{L,Q}$ is transitive: finite $\slpr$ transformations can be used to map any unbound geodesic in $\Gamma_{L,Q}$ to any other. 

In Kerr, the $\slpr$-invariant locus in phase space is the photon shell.
As in Schwarzschild, the large finite dilation \eqref{eq:FiniteDilation} scales down $\th$ and scales up $T$, pushing any trajectory asymptotically onto the homoclinic orbits at large times.
The dimensionless radius of the point of closest approach of the orbit $r=\tr\pa{1+R_{\rm min}}$ becomes 
\begin{align}
	R_{\rm min}^2=-\frac{\pa{\tr+3M}\pa{\tr-M}\Delta(\tr)}{2\tilde{H}\tr\br{\tr\pa{\tr^2-3M\tr+3M^2}-a^2M}}\th\;,
\end{align}
so \eqref{eq:RadialDeviation} still holds.
In Kerr, it is convenient to characterize the approach to criticality by the (fractional) half-orbit number $n_{\rm orb}$ (not to be confused with the QNM overtone number) which diverges as $\alpha\to\infty$ like the inverse power of the Lyapunov exponent \eqref{eq:KerrLyp}
\begin{align}
	\label{eq:half-orbit dilation}
	\pd_\alpha n_{\rm orb}=\frac{1}{\gamma}\;.
\end{align}

\begin{figure}[htp!]
	\centering
	\includegraphics[width=.49\textwidth]{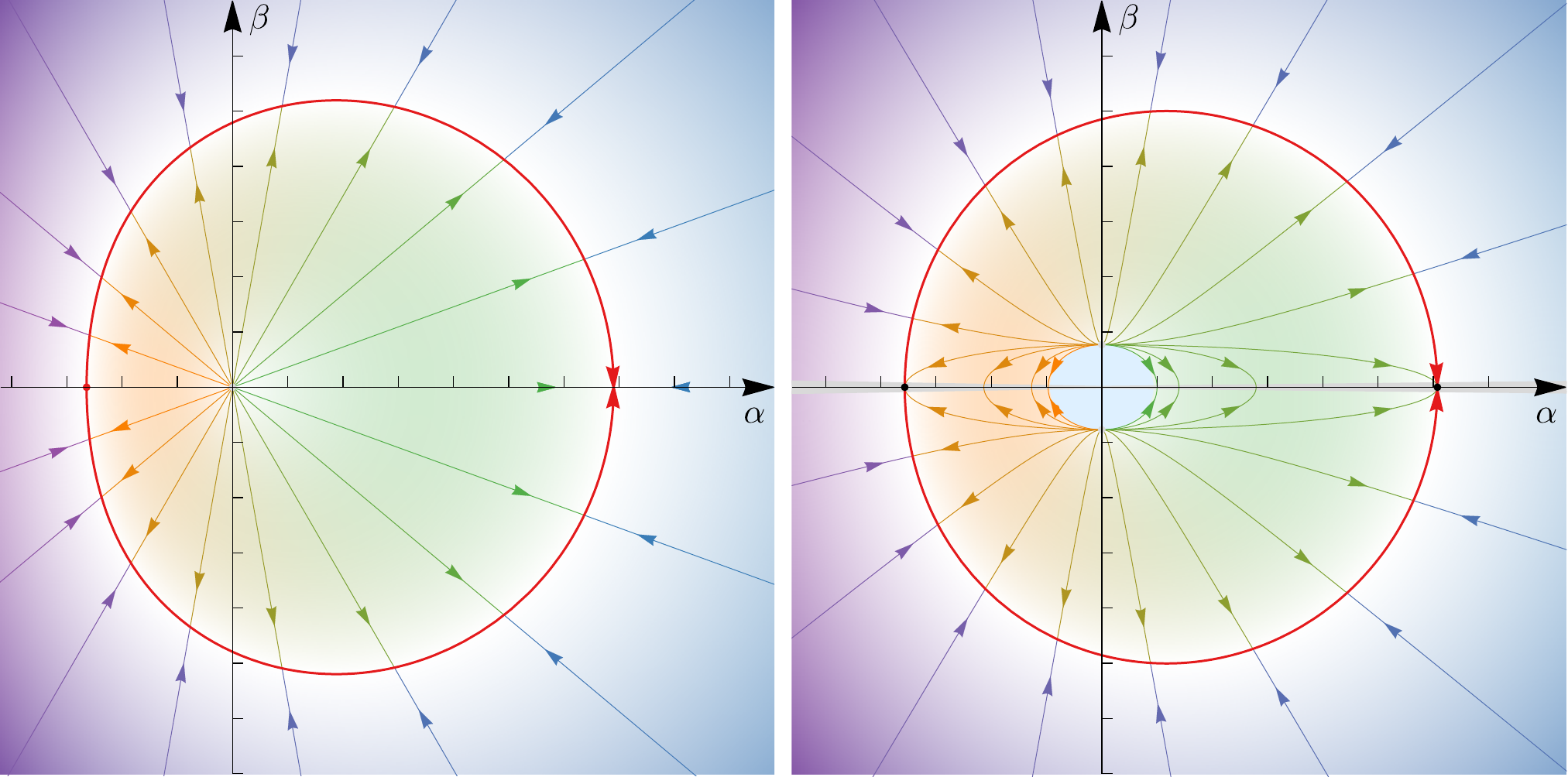}\quad
    \includegraphics[width=.49\textwidth]{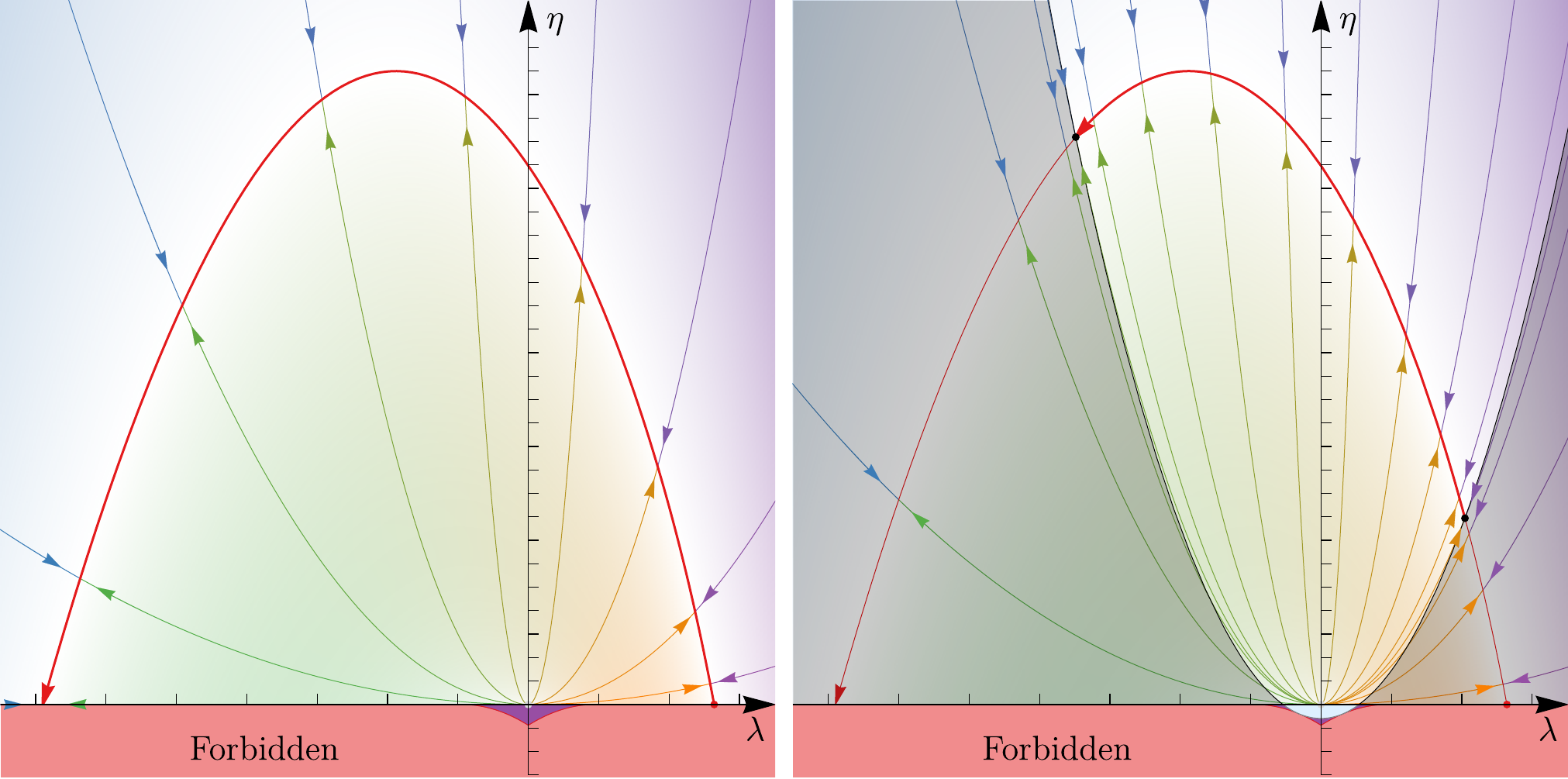}
	\caption{
	Action of $\slpr$ dilations on the image plane of an observer at a large distance from a Kerr black hole, located in the equatorial plane (first panel) or at an inclination of $60^\circ$ (second panel).
	The critical curve (red) corresponds to photons with $\th=0$ that asymptote to bound photon orbits at $\tr$.
	Photons in its interior have $\th>0$ and are captured by the black hole, while those in its exterior have $\th<0$ and are deflected back to null infinity.
	If the photon energy (color) $E$ is fixed, then every choice of $(L,Q)$ defines a point $(\alpha,\beta)$ with coordinates fixed by $\lambda=\frac{L}{E}$ and $\eta=\frac{Q}{E^2}$.
	Otherwise, every $(L,Q)$ defines a ray, with each ray corresponding to the equivalence class of points $(\alpha,\beta)$ whose conserved quantities $(\lambda,\eta)$ are related by energy rescaling.
	The last two panels show the action of $\slpr$ dilations in the phase space of conserved quantities $(\lambda,\eta)$.
	Only the equatorial observer can see the entire photon shell in the sky (first panel), and therefore all of the $\slpr$ orbits in phase space (all of the third panel).
	A non-equatorial observer sees only the subshell of the full photon shell for which $\beta^2(\tr)>0$.
	For the observer at an inclination of $60^\circ$, this corresponds to the unshaded region in the fourth panel.
	Some $\slpr$ flows cross into the shaded region of phase space, which includes the part of the photon shell that is inaccessible to the observer.
	The corresponding dilation flows on the observer screen (second panel) vanish into the horizontal $\alpha$ axis.
	}
	\label{fig:Kerr}
\end{figure}

For the simple and observationally relevant case of stationary, axisymmetric source rings of fixed polar angle $\theta$, the emissivity is independent of $\phi_s$ and emission time $t_s$.
We can then easily repeat the construction of section \ref{subsec:SchScr} for the Kerr black hole.
Considering only geodesics that connect the (axisymmetric) source ring to the telescope cuts the five-dimensional space $\Gamma_{L,Q}\times S^1$ to an infinite, discrete set of circles $S^1$, which we will again call $\Gamma_{\rm obs}$.
Using \eqref{eq:half-orbit dilation}, we identify the $\slpr$ element 
\begin{align}
	D_0=e^{-\gamma H_0}\;,
\end{align}
already discussed above, which maps $\Gamma_{\rm obs}$ to itself while taking $n_{\rm orb}\to n_{\rm orb}+1$.
The semigroup formed by products of $D_0$ is an emergent discrete scaling symmetry of the photon ring.

\section{Quantum Ruelle resonances = classical Lyapunov exponents}
\label{sec:KerrHologram}

In known examples, the $e^{S_{\rm BH}}$ black hole microstates are described by approximately thermal high-energy states in a lower-dimensional quantum field theory, and the response of the black hole to small perturbations is described by linear response theory in the dual quantum mechanics.
This dictionary maps the quasinormal ringing of the black hole atmosphere to the damped oscillations of a perturbed thermal state as it relaxes towards equilibrium \cite{Horowitz2000,Son2002,Birmingham2002,Birmingham2003,Polchinski2015}.

Operating under the relatively mild assumption that the holographic principle applies to asymptotically flat black holes, we expect that Kerr black holes like M87* can be described by a quantum system that we will refer to as the quantum dual.
This \textit{quantum-mechanical} system, if it exists, is constrained by a number of universal features of \textit{classical} black hole physics.
Absent the ability to derive this dual directly from a microscopic theory of quantum gravity, we can instead attempt to infer its properties indirectly from the bottom-up by throwing objects at the black hole and measuring the universal aspects of its response.
As discussed in sections \ref{sec:Schwarzschild} and \ref{sec:Kerr}, the high-frequency part of the Ruelle spectrum has a universal form when expressed in terms of the geometric data of the photon shell.
Following known examples of the holographic dictionary, these QNMs are interpreted as poles in the real-time (retarded) thermal two-point functions in the quantum dual.
In other words, we {\it assume} that these frequencies characterize the long-time ($\Delta t\equiv t-t'\gg T_H^{-1}$) correlations of operators in the quantum dual, which obey
\begin{align}
	\av{\mathcal{O}_{\ell,m}(t)\mathcal{O}_{\ell,-m}(t')}\sim\sum_n e^{-i\omega_{\ell mn}\Delta t}\;.
\end{align}
The brackets denote a thermal average at the Kerr temperature $T_H$ and angular potential $\Omega_H$ in the dual quantum theory\footnote{For black holes in asymptotically flat space, thermal traces (and the partition function itself) do not converge due to negative specific heats and superradiant instabilities, so this expression must be interpreted with some care.} 
\begin{align}
	\av{X}={\rm Tr}\!\br{e^{-\frac{\omega-m\Omega_H}{T_H}}X}\;.
\end{align}
The explicit form of the Ruelle spectrum \eqref{eq:KerrSpectrum} has some salient features.
In a quantum theory, the integers $\ell$ are presumably cut off before $\ell\sim\frac{M}{M_{\rm Planck}}$ when the real parts of the frequencies reach the Planck scale.
As the (rescaled) momentum $\mu=\frac{m}{\ell}$ around the circle runs from $-1$ to $1$, the (rescaled) frequency $\Omega_R(\mu)$ remains positive and increases monotonically
\begin{align}
	-\frac{1}{\tilde{\lambda}(\tr_+)}
	\le\Omega_R(\mu)
	\le\frac{1}{\tilde{\lambda}(\tr_-)}\;.
\end{align}
The dispersion relation $\Omega_R(\mu)$, although universal in Einstein gravity, is complicated and no proposed dual to the Kerr black hole has been able to account for it.

\section*{Acknowledgements}

This work is supported by the Center of Mathematical Sciences and Applications and the Black Hole Initiative at Harvard University, as well as DOE grant de-sc/0007870.
A.L. gratefully acknowledges Will and Kacie Snellings for their generous support.

\bibliography{kls}

\providecommand{\href}[2]{#2}\begingroup\raggedright\begin{thebibliography}{10}

\bibitem{Wheeler}
J.~A. {Wheeler}, ``{Geometrodynamics and the issue of final state},'' in {\em
  Relativity, groups and topology: Proceedings, 40th Summer School of
  Theoretical Physics}, C.~{DeWitt} and B.~S. {DeWitt}, eds.
\newblock Gordon and Breach, 1964.

\bibitem{Zurek1985}
W.~H. {Zurek} and K.~S. {Thorne}, ``{Statistical mechanical origin of the
  entropy of a rotating, charged black hole},''
  \href{http://dx.doi.org/10.1103/PhysRevLett.54.2171}{{\em \prl} {\bfseries
  54} no.~20, (May, 1985) 2171--2175}.

\bibitem{Thorne1986}
K.~S. {Thorne}, R.~H. {Price}, and D.~A. {MacDonald}, {\em {Black holes: The
  membrane paradigm}}.
\newblock Yale University Press, 1986.

\bibitem{Maldacena1997}
J.~{Maldacena} and A.~{Strominger}, ``{Black hole greybody factors and D-brane
  spectroscopy},'' \href{http://dx.doi.org/10.1103/PhysRevD.55.861}{{\em \prd}
  {\bfseries 55} no.~2, (Jan., 1997) 861--870},
  \href{http://arxiv.org/abs/hep-th/9609026}{{\ttfamily arXiv:hep-th/9609026
  [hep-th]}}.

\bibitem{Penington2020}
G.~{Penington}, ``{Entanglement wedge reconstruction and the information
  paradox},'' \href{http://dx.doi.org/10.1007/JHEP09(2020)002}{{\em \jhep}
  {\bfseries 2020} no.~9, (Sept., 2020) 2},
  \href{http://arxiv.org/abs/1905.08255}{{\ttfamily arXiv:1905.08255
  [hep-th]}}.

\bibitem{Almheiri2019}
A.~{Almheiri}, N.~{Engelhardt}, D.~{Marolf}, and H.~{Maxfield}, ``{The entropy
  of bulk quantum fields and the entanglement wedge of an evaporating black
  hole},'' \href{http://dx.doi.org/10.1007/JHEP12(2019)063}{{\em \jhep}
  {\bfseries 2019} no.~12, (Dec., 2019) 63},
  \href{http://arxiv.org/abs/1905.08762}{{\ttfamily arXiv:1905.08762
  [hep-th]}}.

\bibitem{Almheiri2020}
A.~{Almheiri}, R.~{Mahajan}, J.~{Maldacena}, and Y.~{Zhao}, ``{The Page curve
  of Hawking radiation from semiclassical geometry},''
  \href{http://dx.doi.org/10.1007/JHEP03(2020)149}{{\em \jhep} {\bfseries 2020}
  no.~3, (Mar., 2020) 149}, \href{http://arxiv.org/abs/1908.10996}{{\ttfamily
  arXiv:1908.10996 [hep-th]}}.

\bibitem{EHT2019a}
{Event Horizon Telescope Collaboration}, ``{First M87 Event Horizon Telescope
  Results. I. The Shadow of the Supermassive Black Hole},''
  \href{http://dx.doi.org/10.3847/2041-8213/ab0ec7}{{\em \apjl} {\bfseries 875}
  no.~1, (Apr, 2019) L1}, \href{http://arxiv.org/abs/1906.11238}{{\ttfamily
  arXiv:1906.11238 [astro-ph.GA]}}.

\bibitem{Gralla2019}
S.~E. {Gralla}, D.~E. {Holz}, and R.~M. {Wald}, ``{Black hole shadows, photon
  rings, and lensing rings},''
  \href{http://dx.doi.org/10.1103/PhysRevD.100.024018}{{\em \prd} {\bfseries
  100} no.~2, (July, 2019) 024018},
  \href{http://arxiv.org/abs/1906.00873}{{\ttfamily arXiv:1906.00873
  [astro-ph.HE]}}.

\bibitem{Johnson2020}
M.~D. {Johnson}, A.~{Lupsasca}, A.~{Strominger}, G.~N. {Wong}, S.~{Hadar},
  D.~{Kapec}, R.~{Narayan}, A.~{Chael}, C.~F. {Gammie}, P.~{Galison}, D.~C.~M.
  {Palumbo}, S.~S. {Doeleman}, L.~{Blackburn}, M.~{Wielgus}, D.~W. {Pesce},
  J.~R. {Farah}, and J.~M. {Moran}, ``{Universal interferometric signatures of
  a black hole's photon ring},''
  \href{http://dx.doi.org/10.1126/sciadv.aaz1310}{{\em Science Advances}
  {\bfseries 6} no.~12, (Mar., 2020) eaaz1310},
  \href{http://arxiv.org/abs/1907.04329}{{\ttfamily arXiv:1907.04329
  [astro-ph.IM]}}.

\bibitem{Himwich2020}
E.~{Himwich}, M.~D. {Johnson}, A.~{Lupsasca}, and A.~{Strominger}, ``{Universal
  polarimetric signatures of the black hole photon ring},''
  \href{http://dx.doi.org/10.1103/PhysRevD.101.084020}{{\em \prd} {\bfseries
  101} no.~8, (Apr., 2020) 084020},
  \href{http://arxiv.org/abs/2001.08750}{{\ttfamily arXiv:2001.08750 [gr-qc]}}.

\bibitem{GrallaLupsasca2020a}
S.~E. {Gralla} and A.~{Lupsasca}, ``{Lensing by Kerr black holes},''
  \href{http://dx.doi.org/10.1103/PhysRevD.101.044031}{{\em \prd} {\bfseries
  101} no.~4, (Feb., 2020) 044031},
  \href{http://arxiv.org/abs/1910.12873}{{\ttfamily arXiv:1910.12873 [gr-qc]}}.

\bibitem{Hadar2021}
S.~{Hadar}, M.~D. {Johnson}, A.~{Lupsasca}, and G.~N. {Wong}, ``{Photon ring
  autocorrelations},''
  \href{http://dx.doi.org/10.1103/PhysRevD.103.104038}{{\em \prd} {\bfseries
  103} no.~10, (May, 2021) 104038},
  \href{http://arxiv.org/abs/2010.03683}{{\ttfamily arXiv:2010.03683 [gr-qc]}}.

\bibitem{Gralla2020}
S.~E. {Gralla}, ``{Measuring the shape of a black hole photon ring},''
  \href{http://dx.doi.org/10.1103/PhysRevD.102.044017}{{\em \prd} {\bfseries
  102} no.~4, (Aug., 2020) 044017},
  \href{http://arxiv.org/abs/2005.03856}{{\ttfamily arXiv:2005.03856
  [astro-ph.HE]}}.

\bibitem{GrallaLupsasca2020c}
S.~E. {Gralla} and A.~{Lupsasca}, ``{Observable shape of black hole photon
  rings},'' \href{http://dx.doi.org/10.1103/PhysRevD.102.124003}{{\em \prd}
  {\bfseries 102} no.~12, (Dec., 2020) 124003},
  \href{http://arxiv.org/abs/2007.10336}{{\ttfamily arXiv:2007.10336 [gr-qc]}}.

\bibitem{GLM2020}
S.~E. {Gralla}, A.~{Lupsasca}, and D.~P. {Marrone}, ``{The shape of the black
  hole photon ring: A precise test of strong-field general relativity},''
  \href{http://dx.doi.org/10.1103/PhysRevD.102.124004}{{\em \prd} {\bfseries
  102} no.~12, (Dec., 2020) 124004},
  \href{http://arxiv.org/abs/2008.03879}{{\ttfamily arXiv:2008.03879 [gr-qc]}}.

\bibitem{Chael2021}
A.~{Chael}, M.~D. {Johnson}, and A.~{Lupsasca}, ``{Observing the Inner Shadow
  of a Black Hole: A Direct View of the Event Horizon},''
  \href{http://dx.doi.org/10.3847/1538-4357/ac09ee}{{\em \apj} {\bfseries 918}
  no.~1, (Sept., 2021) 6}, \href{http://arxiv.org/abs/2106.00683}{{\ttfamily
  arXiv:2106.00683 [astro-ph.HE]}}.

\bibitem{Ames1968}
W.~L. {Ames} and K.~S. {Thorne}, ``{The Optical Appearance of a Star that is
  Collapsing Through its Gravitational Radius},''
  \href{http://dx.doi.org/10.1086/149465}{{\em \apj} {\bfseries 151} (Feb.,
  1968) 659}.

\bibitem{Cardoso2021}
V.~{Cardoso}, F.~{Duque}, and A.~{Foschi}, ``{Light ring and the appearance of
  matter accreted by black holes},''
  \href{http://dx.doi.org/10.1103/PhysRevD.103.104044}{{\em \prd} {\bfseries
  103} no.~10, (May, 2021) 104044},
  \href{http://arxiv.org/abs/2102.07784}{{\ttfamily arXiv:2102.07784 [gr-qc]}}.

\bibitem{Chan1997}
J.~S.~F. {Chan} and R.~B. {Mann}, ``{Scalar wave falloff in asymptotically
  anti-de Sitter backgrounds},''
  \href{http://dx.doi.org/10.1103/PhysRevD.55.7546}{{\em \prd} {\bfseries 55}
  no.~12, (June, 1997) 7546--7562},
  \href{http://arxiv.org/abs/gr-qc/9612026}{{\ttfamily arXiv:gr-qc/9612026
  [gr-qc]}}.

\bibitem{Festuccia2008}
G.~{Festuccia} and H.~{Liu}, ``{A Bohr-Sommerfeld quantization formula for
  quasinormal frequencies of AdS black holes},''
  \href{http://dx.doi.org/10.1166/asl.2009.1029}{{\em Advanced Science Letters}
  {\bfseries 2} (2009) 221--235},
  \href{http://arxiv.org/abs/0811.1033}{{\ttfamily arXiv:0811.1033 [gr-qc]}}.

\bibitem{Cardoso2009}
V.~{Cardoso}, A.~S. {Miranda}, E.~{Berti}, H.~{Witek}, and V.~T. {Zanchin},
  ``{Geodesic stability, Lyapunov exponents, and quasinormal modes},''
  \href{http://dx.doi.org/10.1103/PhysRevD.79.064016}{{\em \prd} {\bfseries 79}
  no.~6, (Mar., 2009) 064016}, \href{http://arxiv.org/abs/0812.1806}{{\ttfamily
  arXiv:0812.1806 [hep-th]}}.

\bibitem{Compere2012}
G.~{Comp{\`e}re}, ``{The Kerr/CFT correspondence and its extensions: a
  comprehensive review},''
  \href{http://dx.doi.org/10.1007/s41114-017-0003-2}{{\em Living Reviews in
  Relativity} {\bfseries 15} (Mar., 2012) 11},
  \href{http://arxiv.org/abs/1203.3561}{{\ttfamily arXiv:1203.3561 [hep-th]}}.

\bibitem{Bredberg2011}
I.~{Bredberg}, C.~{Keeler}, V.~{Lysov}, and A.~{Strominger}, ``{Wilsonian
  approach to fluid/gravity duality},''
  \href{http://dx.doi.org/10.1007/JHEP03(2011)141}{{\em \jhep} {\bfseries 2011}
  (Mar., 2011) 141}, \href{http://arxiv.org/abs/1006.1902}{{\ttfamily
  arXiv:1006.1902 [hep-th]}}.

\bibitem{Minwalla2012}
S.~{Minwalla}, V.~E. {Hubeny}, and M.~{Rangamani},
  \href{http://dx.doi.org/10.1142/9789814350525\_0014}{``{The Fluid/Gravity
  Correspondence},''} in {\em String Theory and its Applications - TASI 2010,
  From meV to the Planck Scale}, M.~{Dine}, T.~{Banks}, and S.~{Sachdev}, eds.,
  pp.~817--859.
\newblock Nov., 2012.
\newblock \href{http://arxiv.org/abs/1107.5780}{{\ttfamily arXiv:1107.5780
  [hep-th]}}.

\bibitem{Birmingham2002}
D.~{Birmingham}, I.~{Sachs}, and S.~N. {Solodukhin}, ``{Conformal Field Theory
  Interpretation of Black Hole Quasinormal Modes},''
  \href{http://dx.doi.org/10.1103/PhysRevLett.88.151301}{{\em \prl} {\bfseries
  88} no.~15, (Apr., 2002) 151301},
  \href{http://arxiv.org/abs/hep-th/0112055}{{\ttfamily arXiv:hep-th/0112055
  [hep-th]}}.

\bibitem{Chen2009}
B.~{Chen} and Z.-b. {Xu}, ``{Quasi-normal modes of warped black holes and
  warped AdS/CFT correspondence},''
  \href{http://dx.doi.org/10.1088/1126-6708/2009/11/091}{{\em \jhep} {\bfseries
  2009} no.~11, (Nov., 2009) 091},
  \href{http://arxiv.org/abs/0908.0057}{{\ttfamily arXiv:0908.0057 [hep-th]}}.

\bibitem{Strominger1996}
A.~{Strominger} and C.~{Vafa}, ``{Microscopic origin of the Bekenstein-Hawking
  entropy},'' \href{http://dx.doi.org/10.1016/0370-2693(96)00345-0}{{\em
  Physics Letters B} {\bfseries 379} (Feb., 1996) 99--104},
  \href{http://arxiv.org/abs/hep-th/9601029}{{\ttfamily arXiv:hep-th/9601029
  [hep-th]}}.

\bibitem{Strominger1998}
A.~{Strominger}, ``{Black hole entropy from near-horizon microstates},''
  \href{http://dx.doi.org/10.1088/1126-6708/1998/02/009}{{\em \jhep} {\bfseries
  1998} no.~2, (Feb., 1998) 009},
  \href{http://arxiv.org/abs/hep-th/9712251}{{\ttfamily arXiv:hep-th/9712251
  [hep-th]}}.

\bibitem{Guica2009}
M.~{Guica}, T.~{Hartman}, W.~{Song}, and A.~{Strominger}, ``{The Kerr/CFT
  correspondence},'' \href{http://dx.doi.org/10.1103/PhysRevD.80.124008}{{\em
  \prd} {\bfseries 80} no.~12, (Dec., 2009) 124008},
  \href{http://arxiv.org/abs/0809.4266}{{\ttfamily arXiv:0809.4266 [hep-th]}}.

\bibitem{Subramanyan2021}
V.~{Subramanyan}, S.~S. {Hegde}, S.~{Vishveshwara}, and B.~{Bradlyn},
  ``{Physics of the Inverted Harmonic Oscillator: From the lowest Landau level
  to event horizons},'' \href{http://dx.doi.org/10.1016/j.aop.2021.168470}{{\em
  Annals of Physics} {\bfseries 435} (Dec., 2021) 168470},
  \href{http://arxiv.org/abs/2012.09875}{{\ttfamily arXiv:2012.09875
  [cond-mat.mes-hall]}}.

\bibitem{Raffaelli2022}
B.~{Raffaelli}, ``{Hidden conformal symmetry on the black hole photon
  sphere},'' \href{http://dx.doi.org/10.1007/JHEP03(2022)125}{{\em \jhep}
  {\bfseries 2022} no.~3, (Mar., 2022) 125},
  \href{http://arxiv.org/abs/2112.12543}{{\ttfamily arXiv:2112.12543 [gr-qc]}}.

\bibitem{Goebel1972}
C.~J. {Goebel}, ``{Comments on the ``vibrations'' of a Black Hole.},''
  \href{http://dx.doi.org/10.1086/180898}{{\em \apjl} {\bfseries 172} (Mar.,
  1972) L95}.

\bibitem{Ferrari1984}
V.~{Ferrari} and B.~{Mashhoon}, ``{New approach to the quasinormal modes of a
  black hole},'' \href{http://dx.doi.org/10.1103/PhysRevD.30.295}{{\em \prd}
  {\bfseries 30} no.~2, (July, 1984) 295--304}.

\bibitem{Mashhoon1985}
B.~{Mashhoon}, ``{Stability of charged rotating black holes in the eikonal
  approximation},'' \href{http://dx.doi.org/10.1103/PhysRevD.31.290}{{\em \prd}
  {\bfseries 31} no.~2, (Jan., 1985) 290--293}.

\bibitem{Iyer1987a}
S.~{Iyer} and C.~M. {Will}, ``{Black-hole normal modes: A WKB approach. I.
  Foundations and application of a higher-order WKB analysis of
  potential-barrier scattering},''
  \href{http://dx.doi.org/10.1103/PhysRevD.35.3621}{{\em \prd} {\bfseries 35}
  no.~12, (June, 1987) 3621--3631}.

\bibitem{Iyer1987b}
S.~{Iyer}, ``{Black-hole normal modes: A WKB approach. II. Schwarzschild black
  holes},'' \href{http://dx.doi.org/10.1103/PhysRevD.35.3632}{{\em \prd}
  {\bfseries 35} no.~12, (June, 1987) 3632--3636}.

\bibitem{Seidel1990}
E.~{Seidel} and S.~{Iyer}, ``{Black-hole normal modes: A WKB approach. IV. Kerr
  black holes},'' \href{http://dx.doi.org/10.1103/PhysRevD.41.374}{{\em \prd}
  {\bfseries 41} no.~2, (Jan., 1990) 374--382}.

\bibitem{Decanini2003}
Y.~{D{\'e}canini}, A.~{Folacci}, and B.~{Jensen}, ``{Complex angular momentum
  in black hole physics and quasinormal modes},''
  \href{http://dx.doi.org/10.1103/PhysRevD.67.124017}{{\em \prd} {\bfseries 67}
  no.~12, (June, 2003) 124017},
  \href{http://arxiv.org/abs/gr-qc/0212093}{{\ttfamily arXiv:gr-qc/0212093
  [astro-ph]}}.

\bibitem{Dolan2009}
S.~R. {Dolan} and A.~C. {Ottewill}, ``{On an expansion method for black hole
  quasinormal modes and Regge poles},''
  \href{http://dx.doi.org/10.1088/0264-9381/26/22/225003}{{\em Classical and
  Quantum Gravity} {\bfseries 26} no.~22, (Nov., 2009) 225003},
  \href{http://arxiv.org/abs/0908.0329}{{\ttfamily arXiv:0908.0329 [gr-qc]}}.

\bibitem{Dolan2010}
S.~R. {Dolan}, ``{Quasinormal mode spectrum of a Kerr black hole in the eikonal
  limit},'' \href{http://dx.doi.org/10.1103/PhysRevD.82.104003}{{\em \prd}
  {\bfseries 82} no.~10, (Nov., 2010) 104003},
  \href{http://arxiv.org/abs/1007.5097}{{\ttfamily arXiv:1007.5097 [gr-qc]}}.

\bibitem{Dolan2011}
S.~R. {Dolan} and A.~C. {Ottewill}, ``{Wave propagation and quasinormal mode
  excitation on Schwarzschild spacetime},''
  \href{http://dx.doi.org/10.1103/PhysRevD.84.104002}{{\em \prd} {\bfseries 84}
  no.~10, (Nov., 2011) 104002},
  \href{http://arxiv.org/abs/1106.4318}{{\ttfamily arXiv:1106.4318 [gr-qc]}}.

\bibitem{Decanini2010}
Y.~{D{\'e}canini} and A.~{Folacci}, ``{Regge poles of the Schwarzschild black
  hole: A WKB approach},''
  \href{http://dx.doi.org/10.1103/PhysRevD.81.024031}{{\em \prd} {\bfseries 81}
  no.~2, (Jan., 2010) 024031}, \href{http://arxiv.org/abs/0906.2601}{{\ttfamily
  arXiv:0906.2601 [gr-qc]}}.

\bibitem{Yang2012}
H.~{Yang}, D.~A. {Nichols}, F.~{Zhang}, A.~{Zimmerman}, Z.~{Zhang}, and
  Y.~{Chen}, ``{Quasinormal-mode spectrum of Kerr black holes and its geometric
  interpretation},'' \href{http://dx.doi.org/10.1103/PhysRevD.86.104006}{{\em
  \prd} {\bfseries 86} no.~10, (Nov., 2012) 104006},
  \href{http://arxiv.org/abs/1207.4253}{{\ttfamily arXiv:1207.4253 [gr-qc]}}.

\bibitem{GrallaLupsasca2020b}
S.~E. {Gralla} and A.~{Lupsasca}, ``{Null geodesics of the Kerr exterior},''
  \href{http://dx.doi.org/10.1103/PhysRevD.101.044032}{{\em \prd} {\bfseries
  101} no.~4, (Feb., 2020) 044032},
  \href{http://arxiv.org/abs/1910.12881}{{\ttfamily arXiv:1910.12881 [gr-qc]}}.

\bibitem{Gates2020}
D.~E.~A. {Gates}, S.~{Hadar}, and A.~{Lupsasca}, ``{Maximum observable
  blueshift from circular equatorial Kerr orbiters},''
  \href{http://dx.doi.org/10.1103/PhysRevD.102.104041}{{\em \prd} {\bfseries
  102} no.~10, (Nov., 2020) 104041},
  \href{http://arxiv.org/abs/2009.03310}{{\ttfamily arXiv:2009.03310 [gr-qc]}}.

\bibitem{Yang2014}
H.~{Yang}, F.~{Zhang}, A.~{Zimmerman}, and Y.~{Chen}, ``{Scalar Green function
  of the Kerr spacetime},''
  \href{http://dx.doi.org/10.1103/PhysRevD.89.064014}{{\em \prd} {\bfseries 89}
  no.~6, (Mar., 2014) 064014}, \href{http://arxiv.org/abs/1311.3380}{{\ttfamily
  arXiv:1311.3380 [gr-qc]}}.

\bibitem{Kapec2020}
D.~{Kapec} and A.~{Lupsasca}, ``{Particle motion near high-spin black holes},''
  \href{http://dx.doi.org/10.1088/1361-6382/ab519e}{{\em Classical and Quantum
  Gravity} {\bfseries 37} no.~1, (Jan., 2020) 015006},
  \href{http://arxiv.org/abs/1905.11406}{{\ttfamily arXiv:1905.11406
  [hep-th]}}.

\bibitem{Horowitz2000}
G.~T. {Horowitz} and V.~E. {Hubeny}, ``{Quasinormal modes of AdS black holes
  and the approach to thermal equilibrium},''
  \href{http://dx.doi.org/10.1103/PhysRevD.62.024027}{{\em \prd} {\bfseries 62}
  no.~2, (July, 2000) 024027},
  \href{http://arxiv.org/abs/hep-th/9909056}{{\ttfamily arXiv:hep-th/9909056
  [hep-th]}}.

\bibitem{Son2002}
D.~T. {Son} and A.~O. {Starinets}, ``{Minkowski-space correlators in AdS/CFT
  correspondence: recipe and applications},''
  \href{http://dx.doi.org/10.1088/1126-6708/2002/09/042}{{\em \jhep} {\bfseries
  2002} no.~9, (Sept., 2002) 042},
  \href{http://arxiv.org/abs/hep-th/0205051}{{\ttfamily arXiv:hep-th/0205051
  [hep-th]}}.

\bibitem{Birmingham2003}
D.~{Birmingham}, I.~{Sachs}, and S.~N. {Solodukhin}, ``{Relaxation in conformal
  field theory, Hawking-Page transition, and quasinormal or normal modes},''
  \href{http://dx.doi.org/10.1103/PhysRevD.67.104026}{{\em \prd} {\bfseries 67}
  no.~10, (May, 2003) 104026},
  \href{http://arxiv.org/abs/hep-th/0212308}{{\ttfamily arXiv:hep-th/0212308
  [hep-th]}}.

\bibitem{Polchinski2015}
J.~{Polchinski}, ``{Chaos in the black hole S-matrix},'' {\em arXiv e-prints}
  (May, 2015) arXiv:1505.08108,
  \href{http://arxiv.org/abs/1505.08108}{{\ttfamily arXiv:1505.08108
  [hep-th]}}.

\end{thebibliography}\endgroup
\bibliographystyle{utphys}

\end{document}